\documentclass[preprint,superscriptaddress]{revtex4}
\usepackage[dvips]{graphicx}
\usepackage{setspace}
\usepackage{amssymb,amsfonts,amsmath}

\begin{document}
\begin{center}
{\large \bf
The Effects of Stator Compliance, Backs Steps, Temperature, and Clockwise Rotation on the Torque-Speed Curve of Bacterial Flagellar Motor.
\\}
\vspace{0.5cm}
Giovanni Meacci$^{1}$, Ganhui Lan, and Yuhai Tu$^{2}$\\
\vspace{0.3cm} IBM T. J. Watson Research Center\\ P.O. Box 218, Yorktown
Heights, NY 10598\\
\vspace{0.3cm}
$^1$\small{Present address: Departments of Biological Sciences and Chemical Engineering,\\ Columbia University, New York, NY 10027}\\
$^2$\small{Corresponding author (E-mail: yuhai@us.ibm.com; Phone:+1 914 9452762)}\\
\end{center}
\vspace{0.3cm}

\begin{center}
{\bf Abstract}
\end{center}
Rotation of a single bacterial flagellar motor is powered by multiple
stators tethered to the cell wall. In a ``power-stroke" model the observed independence of the speed at low load
on the number of stators is explained by a torque-dependent stepping mechanism
independent of the strength of the stator tethering spring.
On the other hand, in models that depend solely on the stator spring to explain the observed behavior, exceedingly small stator spring constants are required.
To study the dynamics of the motor driven by external forces (such as those exerted by an optical tweezer), back-stepping is introduced when stators are driven far out of equilibrium.
Our model with back-stepping reproduces the observed absence of a barrier to backward rotation, as well the behaviors in the
high-speed negative-torque regime. Recently measured temperature dependence of the motor speed near
zero load (Yuan \& Berg 2010 Biophys J) is explained quantitatively by the thermally activated stepping rates in our model.
Finally, we suggest that the general mechanical properties of all molecular motors (linear and rotary), characterized by their
force(torque)-speed curve, can be determined by their power-stroke potentials and the dependence of the stepping rates on the mechanical state of the motor (force or speed). The torque-speed curve for the clockwise rotating flagellar motor has been observed for the first time recently (Yuan et al. 2010 PNAS). Its quasi-linear behavior is quantitatively reproduced by our model. In particular, we show that concave and convex shapes of the torque-speed curve can be achieved by changing the interaction potential from linear to quadratic form. We also show that reversing the stepping rate dependence on force (torque) can lead to non-monotonicity in the speed-load dependency.

\section*{INTRODUCTION}
Like all other motile bacteria \emph{Escherichia coli} continuously explores its environment in pursuit of favorable conditions for its survival. {\it E. coli} cell's swimming motion is propelled by the rotations of several flagella, each driven by a rotary flagellar motor attached to the cell body. The cell changes its direction of motion by switching the rotational direction of its flagellar motors \cite{Berg_Anderson73,Berg03,Sowa_Berry08}. When all flagellar filaments are driven to rotate counterclockwise (CCW), they form a bundle that propels the cell forward in a smooth swimming (``run") mode. When at least one of the flagella rotates clockwise (CW), the coherent bundle disassembles and the cell tumbles as a result. The cell's signaling pathway enables it to measure and compare environmental changes over time. Positive changes -increases of chemoattractant - prolong the duration of the smooth swimming motions, so that the random walk of the cell is biased towards a preferred direction, i.e., the direction with higher attractant concentrations.

Each flagellar filament ($\sim 10\mu$m long) is driven by a rotatory motor embedded in the cell wall, with an angular speed of the order of $100$ Hz \cite{Berg03}. A single motor is composed of one rotor, a circular ring structure roughly $45$nm in diameter \cite{TFXD06}, and several stator units anchored to the rigid peptidoglycan cell wall. The rotation of the rotor drives the flagellar filament through a short hook. The rotor contains a ring of $\sim 26$ FliG proteins and each stator has four copies of proteins MotA and two copies of proteins MotB, forming two proton-conducting transmembrane channels. A flow of protons, driven by the electrochemical gradients across the channels, causes conformational changes of the stator proteins that generate force on the rotor through electrostatic interaction between MotA and protein FliG \cite{KB01}. In some alkalophilic and marine \emph{Vibrio} species of bacteria, sodium ions take the place of protons. The work per unit charge that a proton can do in crossing the cytoplasmic membrane through the proton channel is called the ``proton-motive force'' (\emph{pmf}). Which consists of two terms: a transmembrane voltage $\Delta\psi$ and an entropic contribution due to the ion concentration difference across the cytoplasmic membrane proportional to $\Delta$pH$=$log$([\textrm{H}^{+}_{\textrm{Periplasm}}]/[\textrm{H}^{+}_{\textrm{Cytoplasm}}])$:
\begin{eqnarray}
\Delta \mu \equiv PMF = \Delta \psi + 2.303 \frac{k_B T}{e} \Delta pH.\label{pmf}
\end{eqnarray}
In the case of a sodium driven motor, $\Delta pH$ is replaced by the sodium ion concentration term $\Delta pNa = - log ([Na^+]_{\textrm{Cytoplasm}}/[Na^+]_{\textrm{Periplasm}}) $.

The duty ratio of the flagellar motor is close to unity \cite{RBB00}, which means that at any given time, at least one stator is engaged with the rotor. Biochemical and structural studies indicate that the motor torque is generated by stator conformational changes as a consequence of ion binding and unbinding to the negatively charged D32 residue on the MotB helices (D24 on PomB for the Na$^+$ motor). Only few portions of the atomic structure of the rotor are known \cite{LWBH99,BHB02,BMJHB05}, and details of the stator-rotor interaction remain unclear due to lack of structural information. Presumably, the passage of (one or a few) protons switches the stator to be engaged with the next FliG monomer on the FliG ring along the direction of rotation, stretching the link between the stator and the rotor. The subsequent relaxation process rotates the rotor and the attached load towards the new equilibrium position. This can give rise to a step-like motion, characterized by advances of the rotor followed by waiting periods, as has been demonstrated by direct observation \cite{SRLYHIB05} for a sodium-powered motor at very low ``sodium-motive-force''.
However, a general understanding of the stepping dynamics of a single flagellar motor is still lacking.

The torque-speed dependence is the key biophysical property that characterizes the motor \cite{Berg03,CB00a,SHHI03}. For \emph{E. coli} at room temperature under physiologically relevant conditions, the maximum angular velocity is $\approx 300$Hz and the estimated maximum torque ranges from $1400$~pN$\cdot$nm \cite{Fahrner03,RLCLAB06} to $4700$~pN$\cdot$nm \cite{BB97}. The measured torque-speed curve for bacterial flagellar motor in the CCW rotational state shows two distinctive regimes: i) a plateau at high load-low speed and ii) a fast quasi-linear decay at low load-high speed. At the transition between the two regimes the angular velocity reaches a large fraction ($\approx 60 \%$) of the maximum velocity, the torque falls by only $\sim10\%$ of its maximum value, $\tau_{max}$, and the torque-speed curve exhibits a characteristic concave shape, like a ``knee" with the leg pulling forward. The general mechanism responsible for the shape (concavity) of the torque-speed curve remains to be explored, particularly in connection with other molecular motors, such as kinesin and myosin-V, where a convex, at high load, and a concave, at low load, force-speed curve was observed \cite{block1999kinesin,block2000force}. In addition, new experimental results \cite{Yuan10} for the CW rotation case show a different torque-speed curve shape respect to the CCW case.

Another important characteristic of the torque-speed curve is the absence of a barrier to backward rotations. Electrorotation \cite{BB99} and optical tweezer \cite{BB97} experiments show almost a constant value for the torque near zero speed, when the motor is driven backwards. The torque does not increase and remains at a constant value in a range up to 100Hz in both directions of the rotational speed. The mechanism for the absence of a backward rotation barrier is unclear.

Several mathematical models \cite{L88,B93,Meister89,BPGEVPB99,WC00,S03} have been proposed to explain various aspects of the observed torque-speed characteristics based on assumptions about details of the electrostatic interaction between the stators and the rotor. More recent works \cite{XBBO06,Bai_et_al09,vanAlbada2009,MYW09} use a general approach independent of the details of the microscopic interactions.
The model introduced in \cite{XBBO06} can reproduce the observed torque-speed curve characteristics.
However, it predicts the maximum velocities depend inversely on the number of stators, Supporting Text in \cite{XBBO06}, which is inconsistent with the observed behavior at low load, where a recent experiment \cite{Yuan_Berg08} shows that near zero load the velocity of the motor is independent of the number of stators. Recently, we have introduced a model of the motor in which the stepping probability of a stator depends on its interaction with the rotor \cite{Meacci_Tu09}. We show that under the general assumption that the stepping rate is higher when the stator is generating torque in the ``wrong" direction, i.e., opposite to the rotational direction of the rotor, the zero-load speed at zero load is independent of the number of the stators. An alternative explanation \cite{Yuan_Berg08,Bai_et_al09} was put forth by introducing a spring between the stator and its anchoring point at the rigid cell wall. The idea to model the stator as a spring anchored to the cell wall, was introduced for the first time in \cite{KB83}. The flexible stator location enabled by a soft spring can damp the conflicting interactions between stators at low load. However, in this explanation is missing a precise study of the chemical and mechanical parameter values, such as the spring stiffness, in order to verify that not only the zero-load speed independence on the stator number is obtained, but also that the motor show a realistic behavior in all of its parts as well that these parameters values are reasonable and cover a sufficient large interval.

Here, we study the effects of the stator spring systematically. We show that the presence of the stator spring does not interfere with the torque-dependent stepping mechanism. On the other hand, in models that depend solely on the the soft stator spring to explain the observed behavior, only a narrow range of small torsional constant spring $\kappa_S$ can reproduce the experimental observation with reasonable stator position fluctuation and displacement.

As described before, the flagellar motor can be driven out of its normal operational range by external forces. The measured torque-speed relationship in these forced regime can inform us about the nature of the stator-rotor interaction and the stator stepping mechanism and test different model predictions. For the negative velocity regime (driven backwards by external forces, such as a optical tweezer), the torque-speed curve is observed to be continuous without any barrier, which is inconsistent with the predictions from models that use periodic potential to describe the stator-rotor interaction \cite{XBBO06,vanAlbada2009,Bai_et_al09} (see for instance Figure 4B in \cite{XBBO06} for the potential shape and Supplementary Information for a detailed discussion).
Here, we extend our model to introduce the possibility of back-stepping which becomes significant as the system is driven far away from the stator-rotor equilibrium. This extension allows us to analyze the negative speed regime; and our model reproduces the absence of a barrier to backwards rotation.
We also explore the negative torque region where the motor is driven forward by the external force beyond its intrinsic zero-load speed.  There, our model again reproduces the experimentally observed behavior \cite{Turner96}, i.e., the torque first drops linearly below zero before  leveling off at a maximum negative torque, comparable in magnitude to the positive maximum torque generated by the stator.

Besides {\it pmf}, the motor dynamics is also affected by the temperature. We use our model to study the temperature effects by assuming a thermally activated stepping rate and compare the model results with the recently measured temperature dependence of the speed at near zero load. The resulting activation energy informs us about the possible gating mechanism for the flagellar motor stepping. Finally, our simple model can be used to address the general question of how the form of the potential describing the interaction between stator and rotor determines the torque-speed curve, in particular its general shape. For example, while the quasi-linear rotor-stator potential results to a concave torque-speed curve, a parabolic potential gives rise to a torque-speed curves with convex or nearly linear shape, similar to the one observed in the CW rotational case \cite{Yuan10}.
Force-velocity curves observed in linear motors such as kinesin-1 and myosin-V also show convex shapes at high load and concave shapes at low load \cite{block1999kinesin,block2000force,lan_sun2005a}
This general relationship provides us a useful tool to predict, either the potential shape when the torque-speed/force-velocity curve is known, or the torque-speed/force-velocity curve when the potential shape is known.

\section*{RESULTS}

\subsection*{The power-stroke model of bacterial flagellar motor }

We use the general model framework in our previous work \cite{Meacci_Tu09} with several important new ingredients. We now include the probability of stator back-stepping and the stator position is anchored to the immobile cell wall through a spring. We first briefly recount the main features of the general model. In Figure 1A, a rotor and a stator are shown with the interacting potential. The two springs represent the load-rotor interaction and the elastic link between the stator units and the peptidoglycan cell wall. Each stator has two force-generating subunits. The interaction potential between the rotor and one of the two stator subunits is represented by the red curve.
As explained in \cite{Meacci_Tu09} the two torque-generating subunits of a stator interact with the FliG ring (rotor) in a hand-over-hand fashion, analogous to the way kinesin proteins interact with microtubules \cite{AFB03,YTVS04}.
The forces between the FliG ring and the stators drive the rotation of the rotor.
The forward hand-switch transition corresponds to a shift of the potential energy in the direction of the motor rotation by an angle $\delta_0$ and the subsequent motor motion is governed by this new potential until the next switch of stator subunit occurs.

The dynamics of the rotor angle $\theta$ and the load angle $\theta_L$
can be described by the following Langevin equations:
\begin{eqnarray}
\xi_{R} \frac{d\theta}{dt}=-\frac{\partial}{\partial\theta}\sum_{i=1}^{N}V(\theta-\theta_i^S)- F(\theta-\theta_{L})
+\sqrt{2k_{B}T \xi_{R}} \alpha(t),\label{e1}
\end{eqnarray}
\begin{eqnarray}
\xi_{L} \frac{d\theta_{L}}{dt}= F(\theta-\theta_{L})+\sqrt{2k_{B}T \xi_{L}} \beta(t),\label{e2}
\end{eqnarray}
where $\xi_{R}$ and $\xi_{L}$ are the drag coefficients for the rotor and the load respectively, and $N$ is the total number of stators in the motor.
$V$ is the interaction potential between the rotor and a stator. $V$ depends on the relative angular coordinates $\Delta\theta_{i}=\theta-\theta_{i}^{S}$, where $\theta_{i}^{S}$ is the internal coordinate of the stator $i$, which consists of two parts: $\theta^S_i=\theta_i+\theta_i^c$. $\theta_i$ is the (physical) angular distance of the stator from its stationary anchoring point on the cell wall; $\theta^c_i$ is the internal (chemical) coordinate of the stator.
The load is coupled to the rotor via a nonlinear spring described by a function $F$, which can be determined from the hook spring compliance measurement of ref. \cite{Block_Berg_Nature_1989}.
The last terms in equations (\ref{e1}-\ref{e2}) are stochastic forces acting on the rotor and on the load, with $k_{B}$ the Boltzmann constant, $T$ the absolute temperature. $\alpha(t)$ and $\beta(t)$ represent two independent white noise fluctuations of unity intensity.

The dynamics of the physical coordinate of the stator $\theta_i$ is governed by a Langevin equation:
\begin{eqnarray}
\xi_{S} \frac{d\theta_i}{dt}=-\frac{\partial}{\partial\theta_i^S}V(\theta-\theta_i^S)- \kappa_S \theta_i
+\sqrt{2k_{B}T \xi_{S}} \gamma_i(t),\label{e3}
\end{eqnarray}
where $\xi_{S}$ is the drag coefficients of the stator, assumed to be the same for all stators, $\kappa_S$ is torsional spring constant of each stator relative to the rotor axis, and $\gamma_i(t)$ is again an independent white noise fluctuation of unity intensity. The internal coordinate of the stator is changed by the hand-switching process, each switching event changes $\theta_i^c$ by $\delta_0$, which is half of
the angular periodicity of the FliG ring, $\pi/26$. A forward step (also called a forward jump) increases $\theta_i^c$ by $\delta_0$. The probability of the forward step during the time interval $\Delta t$ is: $P_f(\theta_i^c\rightarrow\theta_i^c+\delta_0)$. The stepping rate $P_f$ is controlled by the external driving force ({\it pmf}), but it also depends on the mechanical state of the stator characterized by the torque between the rotor and the stator, $\tau_i\equiv -V'(\Delta\theta_i)$, which depends on the relative angle $\Delta\theta_i$:
\begin{equation}
P_f(\theta_i^c\rightarrow\theta_i^c+\delta_0)=R_f(\tau_i)\Delta t=k_f(\Delta\theta_i)\Delta t.
\label{e4}
\end{equation}
In general, we assume $R_f(\tau_i)$ to be a decreasing function of $\tau_i$, e.g., the forward stepping rate is assumed to be higher when the stator generates torque opposite to the rotor rotation \cite{Meacci_Tu09}.

In this paper, we introduce the backward stepping probability $P_b(\theta_i^c\rightarrow\theta_i^c-\delta_0)$, which also depends on the relative angle $\Delta\theta_i$:
\begin{equation}
P_b(\theta_i^c\rightarrow\theta_i^c-\delta_0)=R_b(\tau_i)\Delta t=k_b(\Delta\theta_i)\Delta t.
\label{e3b}
\end{equation}
Backward stepping happens when the stator has overreached on the FliG ring with large value of $\Delta\theta_i$ (compared with $\delta_0$), $P_b$ remains near zero at $|\Delta \theta|_i\le \delta_0$ and increases as $\Delta\theta_i$ increases beyond $\delta_0$. Although back stepping only occurs relatively rarely under normal motor operating conditions as the rotation of the rotor prevents very large values of $\Delta\theta_i$, it becomes more frequent if the rotor is being driven backwards by external forces, which is one of the focuses of our paper.

For the stator-rotor interaction potential $V$, we have considered different functional forms, starting (for simplicity) with the linear $V$-shaped function: $V(\Delta\theta)=\tau_0|\Delta\theta|$, where the torque from a single stator is $\tau_0$ with its sign depending on whether the stator is pulling ($\Delta\theta<0$) or dragging ($\Delta\theta>0$). Correspondingly, the stator forward stepping rate depends on the sign of the force: $k_f(\Delta\theta<-\delta_c)=0$, $k_f(-\delta_c<\Delta\theta<0)=k_+$, $k_f(\Delta\theta>0)=k_-(>k_+)$, with
a cutoff angle $\delta_c$ introduced to prevent run-away stators. The form of the backward stepping rate
will be given in the following sections.
Quantitatively, we use $\tau_0=505$pN$\cdot$nm, $\xi_R=0.02$pN$\cdot$nm$\cdot$s$\cdot$rad$^{-2}$, $\xi_S=0.004$pN$\cdot$nm$\cdot$s$\cdot$rad$^{-2}$, $k_{+}=12000$s$^{-1}$, $k_{-} =2 k_{+}$, $\delta_c=\delta_0$ in this paper unless otherwise stated. The load $\xi_L$ varies from $0.02-50$pN$\cdot$nm$\cdot$s$\cdot$rad$^{-1}$. Other forms of $V$, such as semi-parabolic and pure parabolic potentials will also be studied, with the emphasis on understanding the general relationship between the form of $V$ and the corresponding torque-speed curve.

\subsection*{Independence of the maximum speed on the stator number and the effects of stator spring}

In our previous work \cite{Meacci_Tu09}, it was shown that the maximum rotation speed (at near zero load) is independent of the number of stators under the general condition that the stators are more likely to step forward when they generate backward torque, i.e., $r(=k_+/k_{-})\le 1$. We call this scenario the force-dependent-stepping (FDS) mechanism. The (physical) locations of the stators are fixed in our previous model for simplicity. A more realistic description should take into account that the stator is linked to the rigid peptidoglycan through a $\alpha$-helix of $7$-$8$nm in length as first suggested in \cite{KB83,Yuan_Berg08}. Since the elasticity of $\alpha$-helix is known and its persistence length is about $100$nm \cite{seungho2005elasticity}, the resulting stator spring constant can be then estimated to be at least $500$pN$\cdot$nm$\cdot$rad$^{-1}$, depending on the orientation of the $\alpha$-helix with respect to the rotor. In \cite{Bai_et_al09}, it was shown that the independence of the zero-load speed on the stator number can be obtained by introducing a stator spring with an elastic constant value of $\kappa_S\approx 200$pN$\cdot$nm$\cdot$rad$^{-1}$ or smaller, which is more than 2-fold softer than a $7$-$8$nm $\alpha$-helix spring. In Figure 1B it is shown the dynamics of the rotor and the stator under large load (Supplementary Figure 1 shows also the case at low load) with such a small stator spring constant as used in \cite{Bai_et_al09}.
The angular displacement of the stator from its anchoring point reach as large as $10\delta$, which is 50nm assuming a radius for the rotor of 20nm; and the angular position fluctuation reaches $2\delta$. The large average stator position displacement and its large fluctuation put in doubt this weak-stator-spring (WSS) scenario.

In this section, we systematically investigate the motor dynamics for a wide range of stator spring constants. We emphasize on studying the interplay between the force-dependent-stepping and the weak-stator-spring scenarios.
We show how adding the stator spring does not interfere with the FDS mechanism ($r\le 1$) for a wide range of constant spring values considered. However, very stringent constrains for the stator spring value are required in the absence of the FDS mechanism, i.e., when $r > 1$. Such stringent constrains on the stator spring constant result from two opposing requirements that need to be satisfied by the system. The first requirement is that the stator displacement and the related fluctuation should realistically be smaller than a few $\delta$ \cite{note1}, which sets a lower limit for the spring constant, $k_{S}^{min}$, independent on r. The second requirement is the independence of the zero-load speed on the stator number, which requires the spring to be soft and therefore sets a maximum value for the spring constant, $k_{S}^{max}$, which depends on r. For $r > 1$, these two requirements select a vanishingly small region of possible $\kappa_S$ values.

To characterize the dependence of the zero-load speed on the number of stators, we define the zero-load speed ratio $\omega_r\equiv \omega_{0}(8)/\omega_{0}(1)$, and the zero-load speed variation $\epsilon_r \equiv |1-\omega_r|$, where $\omega_{0}(1)$ and $\omega_{0}(8)$ are the speed at near-zero load for number of stator $N=1$ and $N=8$ respectively. In this case $\omega_{0}(1)= (1 \pm \epsilon_r ) \cdot \omega_{0}(8)$. From the recent resurrection experiments at low load \cite{Yuan_Berg08}, the overall variation of the measured maximum speeds for all stator numbers is $\epsilon_r \sim 5\%$ in single cell, see for example Figure 1B in \cite{Yuan_Berg08}.
Considering the entire range of speeds variation in a cell population in the same experiment, a conservative estimate for $\epsilon_r$ is roughly 10\%, Figure 2A in \cite{Yuan_Berg08}. In the presence of the FDS mechanism, the zero-load speed variation $\epsilon_r$ is less than $20\%$ for all values of the stator spring constant spring considered when $r\le 1$ (Figure 1C), and the variation $\epsilon_r$ quickly decreases as FDS effects become stronger, and when $r \le 0.5$, $\epsilon_r$ becomes smaller than $10\%$. If instead it is less probable for a stator to step (switch-hand) when it is generating the negative torque, i.e. $r>1$, to which the model studied in \cite{Bai_et_al09} belongs, the zero-load speed variation depends on the stator spring constant $\kappa_S$ and only becomes small for small $\kappa_S$. Consider the case r=2,
$\kappa_S$ should be smaller than 20 pN$\cdot$nm$\cdot$rad$^{-1}$ to have independence of the zero-load speed on the stator number within $10\%$, and equal or smaller than 200 pN$\cdot$nm$\cdot$rad$^{-1}$ to have independence within $15\%$ (three times the value estimated from single cell experiment). In addition, soft stator spring leads to large stator position displacement and fluctuations. Figure 1D shows the stator displacement $\Delta\theta^S$ between a stator and its anchor point, and fluctuations interval Var$(\theta^S)$ around the equilibrium position of the stator (during the waiting phase) as a function of the spring constant at high load for the case N=1 and r=2 at high load. Values less of two $\delta$ correspond to values of $\kappa_S$ bigger than 800pN$\cdot$nm$\cdot$rad$^{-1}$. On the other side $\omega_r$ inside a zero-load speed variation of 10\% needs $\kappa_S$ to be smaller than 20pN$\cdot$nm$\cdot$rad$^{-1}$, which makes a strong argument against WSS's models.
All simulation are done with a value of the stator diffusion constant of $D_S=500rad^2s^{-1}$. Other values considered of $50rad^2s^{-1}$ and $5000rad^2s^{-1}$ show very similar behavior. For instance, Supplementary Figures S1A and B show the ratio $\omega_r$ as a function of $\kappa_S$ for different values of r and with stator diffusion constant $D_S=5000rad^2s^{-1}$ and $D_S=50rad^2s^{-1}$.

A more systematic study of the dependence of the model behavior on the two key parameters $r$ and $\kappa_S$ is summarized in Figure 2.
The FDS condition $r<1$ guarantees the independence of the zero-load speed on the stator number (within the experimental error) for a wide range of stator spring constants. If we add in the criterion that the stator fluctuation cannot be bigger than the available space for stator, defined as half of the average spacing between neighboring stators, which is $22.5^{\circ}$ for $N=8$, we obtain a conservative estimation\cite{note2} for the minimum spring constant of $\sim 488$pN$\cdot$nm$\cdot$rad$^{-1}$. This value is similar to the theoretical estimation of the minimum spring constant for the $\alpha$-helix.

On the other hand, in the absence of the FDS condition, only a small region of the parameters (r, $\kappa_S$) that can lead to both small $\epsilon_r$ and reasonable values of stator displacement and fluctuation. The region of acceptable $\kappa_S$ values shrinks as $r$ increases and disappears after $r$ becomes bigger than $\sim 1.5$ (see Figure 2). As well, increasing $\kappa_S$, the initial interval of acceptable r between 1 and 1.5 decreases fast and disappears before $\kappa_S$ reaches the value of $\sim 8,000$pN$\cdot$nm$\cdot$rad$^{-1}$.

\subsection*{Rotation under external force and the stator's backward steps}

An important characteristic of the torque-speed curve for the flagellar motor is the absence of a barrier to backward rotation. Externally applied torques to drive the motor backwards (at negative speed), or forwards at speeds greater than the zero-load speed, have been used since the beginnings of the 90's to discriminate between different models. In particular thermal ratchet models such the one proposed by Meister \cite{Meister89} predicted that the torque will rise sharply if the motor is driven backward. This behavior was first observed \cite{BT93} but successively it turned out to be an artefact of the experimental procedure \cite{Berry96}. Meanwhile, Washizu at al. \cite{Washizu93} found that motor torque is constant up to about 100Hz for rotation in either direction.

In our previous study \cite{Meacci_Tu09} we did not consider backward stator steps, which are relatively rare under normal operating conditions \cite{SRLYHIB05}. However, we expect the back-jumps to become dominant when the stator is driven backwards into regimes with $\Delta\theta <-\delta_c$, where the forward jumps are prohibited. Here, as described in the Model section of the paper, we study the motor behavior under external driving force by including the backward steps (jumps). Since the landing points of these back-jumps are still on the positive side of the potential with positive torque $\tau_0$, inclusion of back-jumps in our model can naturally explain the actually observed torque continuity near stall when the motor is driven backwards. After the first electro-rotation experiment of Washizu et al. \cite{Washizu93} this behavior was confirmed using optical tweezer, where the torque generated by the motor was the same when made to rotate slowly backward as when allowed to rotate slowly forwards at speeds up to 0.3Hz \cite{BB97}. Successive and further electro-rotation experiments \cite{BB99} showed that there is not barrier to backward rotation at a speed up to 40Hz. In one case, data from one particular cell indicate that the linear range for the torque-speed curve extends up to $>$ 100Hz in either directions, confirming again the result found in ref. \cite{Washizu93}. In both experiments cells were tethered to glass coverslip by single flagellum. The \emph{E. coli} strain used carries a \emph{cheY} deletion and thus rotates flagella exclusively CCW. Such a characteristic of the flagellar motor torque-speed curve has been predicted in simple kinetic analysis, involving only stepping rates modeling \cite{BB99}, but the study of the negative speed regime has not been addressed so far in more general descriptions.
Here we explain and interpret these results in the framework of our simple model.

The equation of motion for the load is now modified by an additional term representing the external torque applied directly to the load, for instance by mean of an optical tweezer,
\begin{eqnarray}
\xi_{L} \frac{d\theta_{L}}{dt}= F(\theta-\theta_{L})-\tau_{ext}(t)+\sqrt{2k_{B}T \xi_{L}} \beta(t),\label{e7}
\end{eqnarray}
where $\tau_{ext}$ is the constant external applied torque. When a cell is tethered to a surface by a single flagellar filament, the motor turns the cell body alternately CW and CCW at a speed around 10Hz \cite{Silverman74}. This is the starting working point of our simulations, i.e. the value of the drag coefficient $\xi_{L}$ corresponding to this value of the speed, roughly 5pN$\cdot$nm$\cdot$s$\cdot$rad$^{-2}$, is kept constant. Then the torque-speed curve is obtained by steadily increasing $\tau_{ext}$ by a fixed amount for each point of the curve, mimicking the increasing of the electric field in electro-rotation experiments as described in \cite{BB99}. At a given value of $\tau_{ext}$ the value of the motor torque is calculated by applying the following torque balance equation:
\begin{eqnarray}
\xi_{L} \omega = \tau_{ext} + \tau.
\end{eqnarray}\\
For the jumping rate we use the following expression for the forward jump of each stator i:
\begin{equation}
k_f(\Delta\theta_i)=k_g+k_{0f} \big[exp( (\tau_{i} \delta\theta_l- E_0 )/ k_B T) / \big(1 + exp( (\tau_i \delta\theta_l-E_0) / k_B T)\big)\big]
\end{equation}
and for the backward steps (jumps):
\begin{equation}
k_b(\Delta\theta_i)=k_{0b} \big[exp( - (\tau_{bi} \delta\theta_l - E_0)/ k_B T) / \big(1 + exp( - (\tau_{bi} \delta\theta_l-E_0) / k_B T) \big)\big],
\end{equation}\\
where $\tau_i=-V'(\Delta\theta_i)$ for each stator i, and the force $\tau_{bi}$ represents the slope of a potential equal to the stator-rotor interaction potential but shifted by 0.75$\delta$ respect to the stator position in direction of negative $\Delta\theta_i$ values, see the inset in Figure 3B. $\delta\theta_l$ and $E_0$ set the characteristic torque $\tau_c=E_0/\delta\theta_l$, $k_{0b}=k_r k_{0f}$, with $k_r$ a numerical pre-factor, and $k_g$ a base constant forward rate. As semi-parabolic potential is used here, see Figure 3B. The result does not change when a V-shaped potential is used.

Figure 3B shows the torque-speed curve with N=1 when an external torque is applied to make the rotor rotate backward and forward. In agreement with the experiments, the motor torque varies linearly with speed up to 100Hz in both direction without any barrier (See Supplementary Information for the case N=8).
It is worth mentioning that including the backward jumping probability does not change the qualitative shape and characteristic of the torque-speed curve in its normal operation range in which back stepping rarely occurs. In particular, the independence of the zero-load speed on the number of stators is still hold (see Supplementary Figure S5 for details).

Another interesting region outside of the normal motor operating range is the super high speed negative torque region when the external torque is applied along the same direction as the motor's natural rotation. In this super high speed regime, as the motor speed increases beyond its native maximum speed, the torque generated by the motor drops below zero linearly with the speed before it levels off at a maximum negative torque as measured by Turner et al \cite{Turner96}. Figure 3D shows that our model reproduces this behavior. The reason why the torque levels off at a negative torque value equal in absolute value to $\tau_0$ can be understood from our model as shown in Figure 3C. As the external torque forces the motor move faster than its natural maximum speed at low load, the stators spend most of their time in the negative torque regime of the rotor-stator potential, where the maximum value of the torque is equal to $-\tau_0$. To the extent that the observed maximum negative torque at super high speed is comparable to the motor's natural maximum positive torque, our model shows that the rotor-stator potential is approximately symmetric.

\subsection*{The temperature effect at low loads and the thermally activated stepping}

Changing temperature affects all the chemical transition rates, in particular the stepping rates. The temperature dependence of $k_{+}$ and $k_{-}$ in our model leads to changes in the knee speed $\omega_n$ and the speed near zero load $\omega_{0}$ without changing the maximum torque at stall, which are consistent with previous experimental observations \cite{BT93,CB00b}. In our previous work we did not investigate the thermal effects in details. Here, we carry out the study of the temperature effects on the torque-speed curve.
The main motivation for such a study is given by a recent experiments \cite{yuan2010thermalisotope}, which extended the previous studies \cite{KB83} of thermal effects on motor rotation at the high-load regime to the near zero load regime. They used nano-gold spheres attached to the hooks of the flagella in mutant \emph{E. coli} cells lacking flagellar filaments to measure speed variations of the motor near zero load within a temperature range from 9 to 37$^\circ$C. They found that the speed changes nearly exponentially with the temperature, and a value of 52 kJ/mol for the activation enthalpy was extracted from the temperature dependence of the speeds near zero load.

For simplicity we consider a V-shaped stator-rotor interaction potential, with the temperature dependence of the stepping rates following the simple Arrhenius law:
\begin{equation}
k_{\pm}=k_{0\pm} \, exp(- \Delta G / R T) ,
\label{e35}
\end{equation}
where R=8.3 JK$^{-1}$mol$^{-1}$ is the universal gas constant,
$\Delta G=52$ kJ/mol is the activation enthalpy, $k_{0-}=1.6\times 10^{13} s^{-1}$,
and $k_{0+}/k_{0-}=0.5$. Figure 4A shows the dependence of speed near zero load on temperature obtained from our model. The agreement with the experimental data is excellent with $\Delta G$ given by the experiments and only one fitting parameter $k_{0-}$. At high load, shown in Figure 4B, the motor torque is independent of temperature because the interaction potential $V$ is independent of the temperature. For the temperature range ($\Delta T/T\sim 10\%$) considered here, $k_{0-}$ and $k_{0+}$ are approximately constant. However, both $k_{0+}$ and $k_{0-}$ depend on temperature through a torque-dependent factor: $exp(- \tau \delta_l /k_B T)$, where $\delta_l$ can be interpreted as the size of the (angular) conformational change of the stator's transition state before stepping. A lower bound for $\delta _l$ can be estimated from the maximum torque $\tau_0$ and the value of $r$: $\delta_l\ge [k_B T/2\tau_0] \ln(1/r)$, which is roughly $0.2-0.5^o$ for $r=0.5$ and $\tau_0=175-500$pN$\cdot$nm.

\subsection*{The general properties of the torque(force)-speed relationship and the special case of the CW rotation in BFM}

The most important and measurable biophysical characteristics of a motor is its force-speed curve (for linear motor) or torque-speed curve (for rotatory motor). What determines the overall shape the torque-speed curve? Two important factors are the detailed forms of the interaction potential and the stepping rate. Here we study the general relationship between the these two key factors and the shape of the torque-speed ($\tau-\omega$) curve. In the case of the V-shaped and the semi-parabolic potentials (V-shaped with a smooth bottom), which we have focused on so far, the $\tau-\omega$ curve is concave, see for instance Figure 6A.

Now consider a pure parabolic potential as shown in the inserts of Figure 5A and B and the following
stepping rates:
\begin{equation}
k_f(\Delta\theta_i)=k_{0f} exp(-\tau_i \delta\theta_l / k_B T),
\label{eq:e31}
\end{equation}
\begin{equation}
k_b(\Delta\theta_i)=k_{0b} exp(\tau_i \delta\theta_l / k_B T),
\label{eq:e32}
\end{equation}
which depends exponentially on the torque $\tau_i(=-V'(\Delta\theta_i))$ for the forward and backward steps with constant $k_{0b}$, $k_{0f}$ and $\delta\theta_l$.
Figure 5A shows the torque-speed curves for different values of $N$. The torque-speed curves are slightly convex, almost linear. Another interesting finding is that for this choice of the stepping rate dependence on torque, the maximum speed at low load increases with $N$. This dependence of $\omega_{0}$ on $N$ is a consequence of the unbounded exponential increase of the forward jumping rate with the negative torque. As the number of stator increases, the positions of the stators in the waiting phase can be pushed into regimes with larger values of negative torque due to the quadratic nature of the potential and henceforth larger jumping rates according to equation (\ref{eq:e31}). Consequently, the larger forward jumping rates in the waiting phase lead to higher maximum speeds for larger $N$.

In order to maintain the independence of $\omega_{0}$ on $N$, the jumping rate $r(\tau_i)$ needs to saturate to a maximum value as $|\tau_i|$ increases. For example, we can use the following expressions for the forward and backward jumping rates:
\begin{equation}
k_f(\Delta\theta_i)=k_{0f} \big[exp( (\tau_i \delta\theta_l- E_0 )/ k_B T) / \big(1 + exp( (\tau_i \delta\theta_l-E_0) / k_B T)\big)\big],
\label{eq:e33}
\end{equation}
\begin{equation}
k_b(\Delta\theta_i)=k_{0b} \big[exp( - (\tau_i \delta\theta_l - E_0)/ k_B T) / \big(1 + exp( - (\tau_i \delta\theta_l-E_0) / k_B T)\big)\big].
\label{eq:e34}
\end{equation}
Here, $E_0$ and $\delta\theta_l$ set the force scale $\tau_c=E_0/\delta\theta_l$ beyond which the forward jumping rate saturates. Choosing $\tau_c \approx 0.1 \tau_0$ the independence of the max speed on the stator number is recovered as shown in Figure 5B. This behavior seems to be independent of the specific values of $\tau_c$ and $\delta\theta_l$. Supplementary Figure S5A shows he case of $\tau_c=0.2 \tau_0$ and Supplementary Figure S5B the torque-speed curves for different values of $\delta\theta_l$.

Within this framework our model is able to generate a torque-speed curve in quantitative agreement with the CW torque-speed curve observed experimentally \cite{Yuan10}. Until very recent it has been assumed that CCW and CW rotation are symmetric and exhibit the same torques and speeds \cite{Blair_Berg_88}. However, Yuan et al. measured the torque-speed relationship for an \emph{E. coli} strain locked in CW rotation and found a quasi-linear curve. As discussed in \cite{Yuan10}, the CW torque-speed relationship can be explained within our model by having $k_+$ close to zero (or much smaller than k$_-$) with a V-shaped interaction potential (see Figure S6a of ref. \cite{Meacci_Tu09}), or by having a parabolic stator-rotor potential. Figure 5C shows an excellent agreement between the experimental data and our simulation of the torque-speed curve with bounded jumping rates.

The stepping-rate ratio $r$ not only affects the dependence of the zero-load speed on the number of stators, it also changes the characteristics of the torque speed dependence at low load. Figure 6A shows the torque-speed curves for two different values of the ratio r: $r=2$ and $r=0.5$. For $r=0.5$, the torque $\tau$ generated by the motor decreases as its speed $\omega$ increases. However, for $r=2$, the slope of torque-speed curve changes its sign after the knee region from negative to positive, giving rise to a non-monotonic dependence of speed on torque (or load). (the same behavior is observed when stator springs are added, see Supplementary Figure S4A and B).

The mechanism for this behavior can be understood within our model. As described in \cite{Meacci_Tu09}, the dynamics of the motor can be characterized by the two timescales: the average moving time $<t_m>$ and the average waiting time $<t_w>$. Their dependence on the motor speed are shown in Figure 6B for both $r=0.5$ and $r=2$.

These two different forms of the torque-speed relationship can be understood by investigating the different dependence of $\langle t_w\rangle$ on the load $\xi_L$ for $r=0.5$ and $r=2$ as shown in the inset of Figure 6B. At a specific value of the load $\xi_L^*$ close to the ``knee", where the torque-speed curves of the two cases ($r=0.5$ and $r=2$) start to diverge, the dependence of the average waiting time $<t_w>$ on $\xi_L$ also becomes qualitatively different. While $<t_w>(r=0.5)$ continues to decrease and eventually level off at roughly 60$\mu$s near zero load, $<t_w>(r=2)$ starts to increase at $\xi_L^*$ to a higher value $\sim 110\mu$s, almost twice the value in the $r=0.5$ case. This is caused by the fact that for $r>1$ it is harder to step (switch-hand) when the stator spends more time in the negative torque regime at higher load because $k_-<k_+$. This increase in waiting time for $\xi_L<\xi_L^*$ leads to the decrease in the motor speed as the load is lowered (below $\xi_L^*$) for $r=2$ as seen in Figure 6A.
In Figure 6C it is shown the sign changing in the torque-speed derivative after the ``knee''. The torque-speed derivative goes from positive to negative when r goes from 1.5 to 0.5, with a critical value $r_c=1$, where the derivative is not defined ($\infty$), that divides the positive from the negative derivative case.

In the model studied in \cite{XBBO06,Bai_et_al09}, the stepping rate has a complicated dependence on the relative angle between the stator and the rotor and in general favors stepping in the positive torque regime, which corresponds to $r>1$ in our model. Indeed, the torque-speed curve shows a change of sign in the torque-speed derivative at around the ``knee", consistent with what we observed for $r=2$ in our model, see for example Figure 2B in \cite{Bai_et_al09}. Experimentally, such a non-monotonic torque-speed dependence has never been observed. Therefore, any realistic model should have $r\le 1$ (or the equivalent), where the FDS mechanism is relevant.

\section*{DISCUSSION}

In this paper, our original mathematical model framework for the rotary flagellar motor \cite{Meacci_Tu09} is extended substantially to incorporate key biological ingredients, such as the stator back-stepping, the stator springs, the temperature dependence of the stepping rate, and the CW rotation. Our model allows us to learn about the relevant microscopic mechano-chemical details of the flagellar motor, such as the power-stroke potential and the stator stepping mechanism, from the ``macroscopic'' measurements, in particular the torque-speed dependence. Summary and discussion of specific findings are given below.

\emph{Backward rotation.} The introduction of a torque-dependent back-stepping rate reproduces the continuity of the experimental torque-speed curve near stall (high load). We proposed that the back-stepping probability is dominant in the far-positive side of the V-shaped potential, where the forward-step is negligible. Since the landing points of these back-jumps are still on the positive side of the potential with the same positive torque $\tau_0$, inclusion of back-jumps in our model naturally explains the observed continuous torque-speed relation across the stalling point, when the motor is driven backwards. The torque-speed relationship in the bacterial flagellar motor is linear up to 100Hz in either direction, which rules out models \cite{XBBO06,Bai_et_al09,vanAlbada2009} that only allow the motor to move backward over a potential barrier. Back steps were also considered in \cite{Bai_et_al09}. However, since there the back stepping rate is only controlled by the {\it pmf}, it does not prevent the motor from climbing up and sometimes crossing the potential barrier. A more detailed discussion of alternative models is given in the Supplementary Information.

\emph{Effects of stator springs.} Thorough analysis of parameter dependence of speed at near zero load shows that adding the stator spring preserves our original model results for all ranges of spring constant values as long as $r\le 1$. However, the stator spring constant needs to be highly restricted when $r>1$. In particular, even a very conservative value for the load-free speed variation $|1-\omega(8)/\omega(1)|=15\%$ would constrain the torsional stator spring constant into a very small range. Furthermore, the weak stator spring required for $r>1$ gives rise to unrealistic large displacements and fluctuations of stator positions, which further limit the validity of the soft stator spring scenario.

\emph{Characteristics of the torque-speed relationship.} Performance under external load generally characterizes all motor proteins, including both rotary motors and linear motors. In our simple model, the torque-speed curve is determined by two factors: the power-stroke potential and the dependence of jumping/stepping rates on the mechanical coordinate. For example, we show here that changing the ratio of stepping rates $r=k_+/k_-$ can change the torque-speed dependence from monotonic to non-monotonic; and changing the potential from linear to quadratic can leads the torque-speed curve to go from concave to convex.

The newly measured torque-speed curve for the CW rotating BFM \cite{Yuan10} can be quantitatively reproduced by our model with a parabolic stator-rotor interaction potential. As shown in our previous work \cite{Meacci_Tu09}, such a liner torque-speed curve can also be obtained with a V-shaped potential by setting the jumping rate $k_+$ equal or near to zero. Is it possible to distinguish between the two scenarios? In the low to medium load regime where the recent experiments have been done, these two mechanisms generate the same quasi-linear torque-speed relationship. However, at high load these two scenarios give rise to opposite torque-speed curvature. Specifically, the parabolic potential shows an increasing of the torque much faster than linear whereas the small $k_+$ scenario leads to a constant torque near stall. A detailed measurement of the torque-speed near stall for the CW rotation will be able to distinguish these two scenarios. The experimental data are fitted better by having a bounded jumping rate, which (as shown in Fig. 5B) leads to the independence of the speed near zero load on the number of stators for the CW rotating motor, just like for the CCW case. This prediction can be checked with resurrection experiments for CW motor at low load.

Finally, we notice that the convex shape of torque-speed curves for the parabolic operating potential are very similar to the force-velocity curves of linear motor kinesin-1 at high load \cite{block1999kinesin,block2000force}, also the collapse at maximum velocity agree with current theoretical model for linear kinesin-1 motor \cite{Kunwar_Gross_Curr_Biology_2008} and for linear myosin motor systems \cite{lan2005muscle}. These similarities suggest that our simple modeling framework may be used to understand/categorize the general force-speed dependence for all the mechanical motor molecules.

\emph{Chemical and mechanical gating of the stepping process.}
Within our model framework, temperature affects flagellar motor operation
through modifying the stepping rates. Such thermally activated process can be formally treated by the transition state theory
\cite{laidler1983TST}. From the value of the free energy barrier $\Delta G$ obtained from the temperature dependence of the stepping rate,
stepping seems to be predominately controlled (gated) by the {\it pmf}. Consequently one would expect that the {\it pmf} dependence of the
motor speed near zero load could have an exponential component. More detailed theoretical models, which consider gating mechanisms where proton
translocation is modeled explicitly by a barrier crossing event, report different behaviors. In \cite{XBBO06,Bai_et_al09} the barrier height
depend only on {\it pmf}, and the speed at low load increases almost linearly with the membrane voltage. In \cite{MYW09} both torque and speed
were predicted to grow sub-linearly with {\it pmf}. None of these predictions can be ruled out by the indirect experiments showing a linear
{\it pmf} dependence in a relatively small range of rotation speed between 80Hz and 270Hz \cite{GB03}. We believe direct measurements of the
{\it pmf} dependence of the motor behavior at near-zero load are needed to elucidate the exact chemical gating mechanism of the stator. Besides
{\it pmf}, our study shows that there must also be a mechanical contribution to $\Delta G$, which accounts for the torque-dependent stepping
rate. Depending on the value of $r$, this mechanical gating energy is small ($\sim k_BT$ for $r=0.5$) in comparison with the ${\it pmf}$, but
its effects are significant as shown in \cite{Meacci_Tu09} and here. A lower bound for the conformational change ($\sim 0.2-0.5^{o}$ in angle
or $\sim 0.8-2$ angstrom in distance) in the transitional state of the stator can be deduced from the mechanical gating energy and the maximum
torque.

The pre-factor $k_{0}$ in the stepping rate in equation (\ref{e35}), often interpreted as the intrinsic molecular fluctuation frequency or the
reaction attempting rate, should be proportional to $\sqrt{f/m}$, where $f$ is the stiffness of the harmonic energy landscape of the reaction
and $m$ is the mass of the reacting molecule. For flagellar motor driven by the proton motive force, the stepping rates, when keeping
temperature constant, should be proportional to $1/\sqrt{m}$. Hence, the intrinsic maximum speed of flagellar motor near zero load should be
scaled by $\sqrt{2}$ when changing the operating environment from water ($H_2O$) to heavy water ($D_2O$). This isotope effect has indeed been
confirmed by recent experiments \cite{yuan2010thermalisotope}.

\section*{Supporting information}
Supplementary information.

\section*{Acknowledgements}
We thank Drs. Junhua Yuan and Howard Berg for many stimulating discussions.



\newpage
\noindent
{\bf Figure Legends}
\vspace{0.5cm}

\noindent
{\bf Figure 1. The motor dynamics with stator springs.} ({\bf A}) A a sketch of the model for the flagellar rotary motor with the load (blue dot), linked to the rotor (green dot) with a spring, and a stator linked to the peptidoglycan cell wall (PG) by a spring. The red line represents a possible potential describing the stator-rotor interaction.
      ({\bf B}) The stator, rotor and load dynamics at high load (N=1, r=2), with $\kappa_S =$200pN$\cdot$nm$\cdot$rad$^-1$ and $D_S=$500rad$^2$/s. The black  solid line shows the stepping dynamics of the stator's chemical coordinate, the red dashed line represents the stator's overall dynamics: internal coordinate jumps plus the physical angular movements. The stretching of the stator spring at steady-state is roughly $10 \delta$. Green and blue dashed lines represent the rotor and load angular coordinates respectively.
      ({\bf C}) $\omega_r=\omega(8)/\omega(1)$ as a function of $\kappa_S$ with $D_{S}=500rad^{2}s^{-1}$. Black dots, red squares, and green diamonds, correspond to r=0.5, r=1, and r=2 respectively. The black dashed line represents the 20\% error line: $\omega_r - 20\%$. The light blue semitransparent area represents a zero-load speed variation of 10\% (as estimated from the experiment).
      ({\bf D}) Spring displacement $\Delta\theta^S$, red squares, and fluctuations Var($\theta^S$), black dots, normalized by $\delta$, during the waiting phase as a function of the spring constant at high load for the case N=1, r=2. Reasonable values on the order of two $\delta$ or less correspond to values of $\kappa_S$ bigger than 800pN$\cdot$nm$\cdot$rad$^{-1}$ (black arrow). On the other side $\omega_r$ inside a zero-load speed variation of 10\% needs $\kappa_S$ to be smaller than 20pN$\cdot$nm$\cdot$rad$^{-1}$ (light blue semitransparent area).
\vspace{1cm}

\noindent
{\bf Figure 2. Phase diagram of the motor behavior in the parameter space.} The spring constant $\kappa_S$ is on the x-axis, and the ratio of stepping rates $r$ on the y-axis. The red line with black circles in the main plot shows the parameter choice to achieve $15\%$ zero-load speed variation ($\epsilon_r=15\%$), model parameters below the red line correspond to $\epsilon_r <15\%$. The inset shows the standard deviation (red dots) of individual stator at different spring constant for the case of N=8, r=0.5, and $\xi_L=0.002$pN$\cdot$nm$\cdot$s$\cdot$rad$^{-2}$ (low load). The allowed space for stator position fluctuation is set as half of the average spacing between neighboring stators, and is $22.5^{\circ}$ for $N=8$ which gives rise to the minimum spring constant to be $\sim 488$pN$\cdot$nm$\cdot$rad$^{-1}$ (green arrow and green triangle in the main plot). The two constrains, the zero load speed and the stator position fluctuation, both control the choice of parameters together. For $r\ge 1.0$, the stator spring constant is highly limited, as shown by the small dark gray region, which diminishes when $r\ge 1.5$. Whereas for $r<1.0$, the allowed stator spring constant, as shown by the green region, has much broader choices with essentially no upper limit.
\vspace{1cm}

\noindent
{\bf Figure 3. Dynamics of the motor driven by external forces.} ({\bf A}) The backward forcing scheme. The rotor (green dot) moving on the rotor-stator potential (red curve) is elastically linked to the load (blue dot), to which a force is applied in direction opposite to the force applied by the stator, given by the slope of the potential.
      ({\bf B}) Torque-speed curve for N=1 near stall. Starting from the black dot, the dashed black and red lines represent the cases when the applied external torque has respectively the opposite and the same sign as the stator generated torque. The inset shows the potential between rotor and stator (dashed green line) and the forward and backward rates are represented by the red and black line respectively.
      ({\bf C}) The forward forcing scheme. The applied force is in the same direction as the rotation of the rotor.
      ({\bf D}) The normalized motor torque versus speed as the motor is driven forward by external forces for N=1, solid black line, and N=8, dashed red line. As the external torque drives the motor rotate faster than its natural maximum speed, the motor generates net negative torques. The motor torque eventually saturates at a minimum with its absolute value roughly the same as the motor's natural maximum positive torque for both N=1 and N=8.
\vspace{1cm}

\noindent
{\bf Figure 4. The temperature dependence of the torque-speed curves.} ({\bf A}) Speed near zero load as a function of temperature. Solid red light shadow line is from numerical simulations of our model. The width of the curve corresponds roughly to the error of the maximum speed. Black dots are experimental data from \cite{yuan2010thermalisotope}. Three points (open circles with error bar) are data from \cite{BT93,CB00a}. ({\bf B}) Torque-speed curves at different temperatures, shown in the inset, with stator number from N=1 to N=8. V-shaped potential is used and the parameters for the stepping rates are: $\Delta G=52 kJ/mol$, ${k}_{0-}=exp(30.4)$.
\vspace{1cm}

\noindent
{\bf Figure 5. The torque-speed curves with parabolic potential.} Different number of stators, from N=1 to N=8, and different stepping rate are used.
      ({\bf A}) Unbounded stepping rate, shown in the inset, are given by equations (\ref{eq:e31}) and (\ref{eq:e32}) with parameter values $k_r=0.01$ and $k_0=12000s^{-1}$. Black dashed-dotted line: backward jumping rate; red solid line: forward jumping rate; green dashed line: parabolic stator-rotor potential.
      ({\bf B}) Bounded stepping rates, shown in the inset, are given by equations (\ref{eq:e33}) and (\ref{eq:e34}) with parameter values $E_c=0.1 \tau_0 \delta \theta_l$, $k_r=0.001$ and $k_0=30000s^{-1}$. Parameter values common to ({\bf A}) and ({\bf B}) are: $\kappa_S$=10$^5$pN$\cdot$nm$\cdot$rad$^{-1}$, $D_S=500rad^2 s^{-1}$, $\delta\theta_l=0.1\delta$. Black dashed-dotted line: backward jumping rate; red solid line: forward jumping rate; green dashed line: parabolic stator-rotor potential.
      ({\bf C}) Normalized torque versus speed for CW rotation. Green lines (error bars) are experimental data taken form \cite{Yuan10}. Gray transparent line simulation of our model, with $k_0=32000s^{-1}$ and N=8. All other parameters have the same value as in Figure 5B. In order to compare the simulations with the experimental published data, the normalization of the gray curve has been done at the speed value corresponding to the speed value of the first experimental data, i.e. roughly 61 Hz, and not at its maximum value at near zero speed. Notice, that the maximum normalized torque value for the experimental data is not equal to 1. In fact, the CW data points in \cite{Yuan10} are normalized respect to the CCW curve, not shown here. See \cite{Yuan10} for details.
\vspace{1cm}

\noindent
{\bf Figure 6. The dependence of torque-speed curve on $r$.} ({\bf A}) The torque-speed ($\tau-\omega$) curves for two different values of r with $N=1$. Red squares r=0.5, black dots r=2.
      ({\bf B}) The average waiting-time $<t_{w}>$ for r=0.5 (black solid line) and r=2 (green dotted-dashed line), and the average moving-time $<t_{m}>$ for r=0.5, red dashed solid line, and r=2, blue dashed line, over 500 revolutions as a function of the rotational speed for N = 1. The inset shows the zoom of the average waiting time $<t_{w}>$ as a function of the the inverse of the load, $\xi_L$, for the two cases: r=2, black dashed line, and r=0.5, red dotted line.
      ({\bf C}) The torque-speed ($\tau-\omega$) curves for 11 different values of r, from 1.5 to 0.5. The red solid line represent the curve corresponding to the critical value $r_c=1$.

\clearpage
\begin{figure}
   \begin{center}
      \includegraphics*[angle=0,width=5.25in]{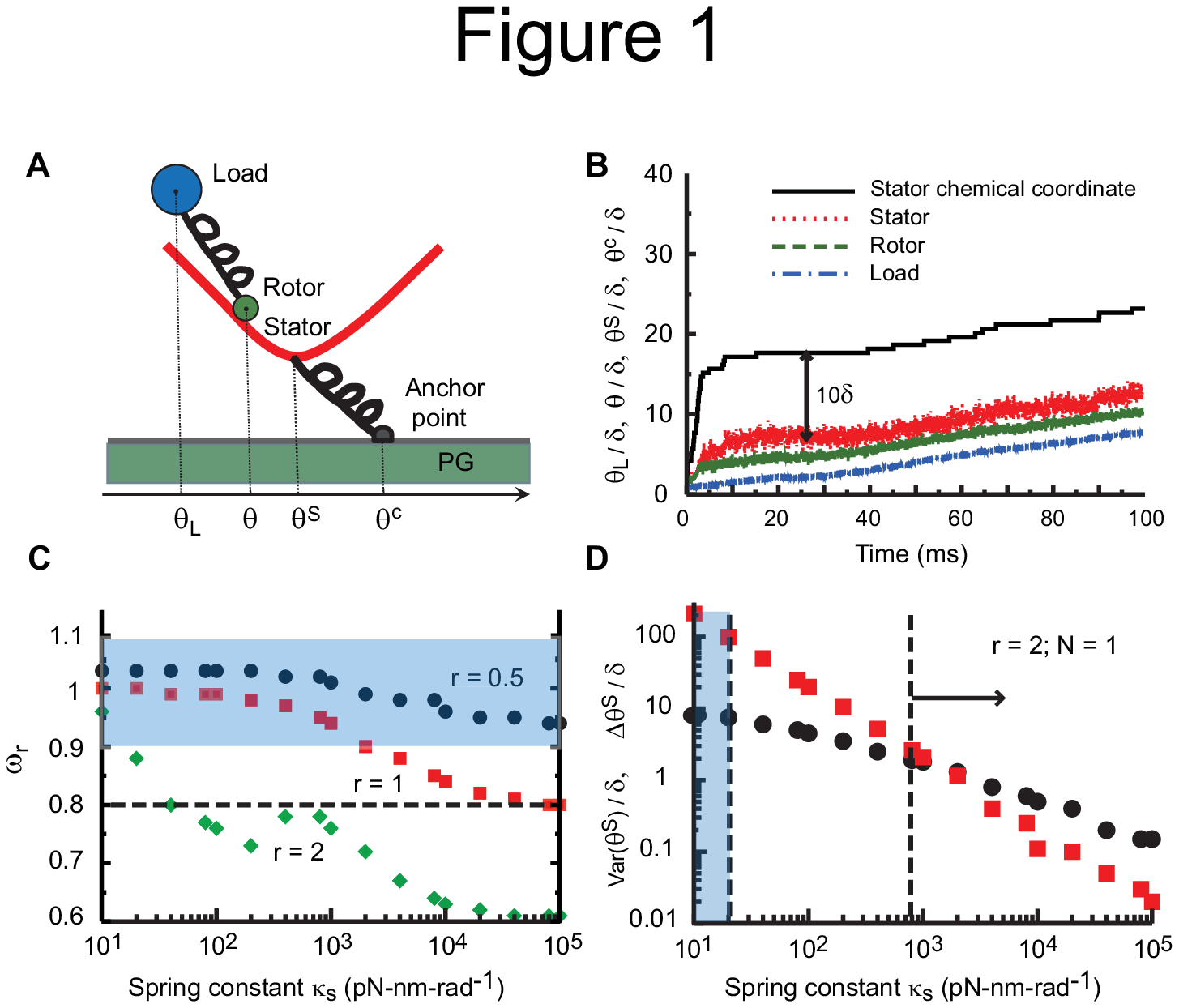}
   \end{center}
\end{figure}

\clearpage
\begin{figure}
   \begin{center}
      \includegraphics*[angle=0,width=5.25in]{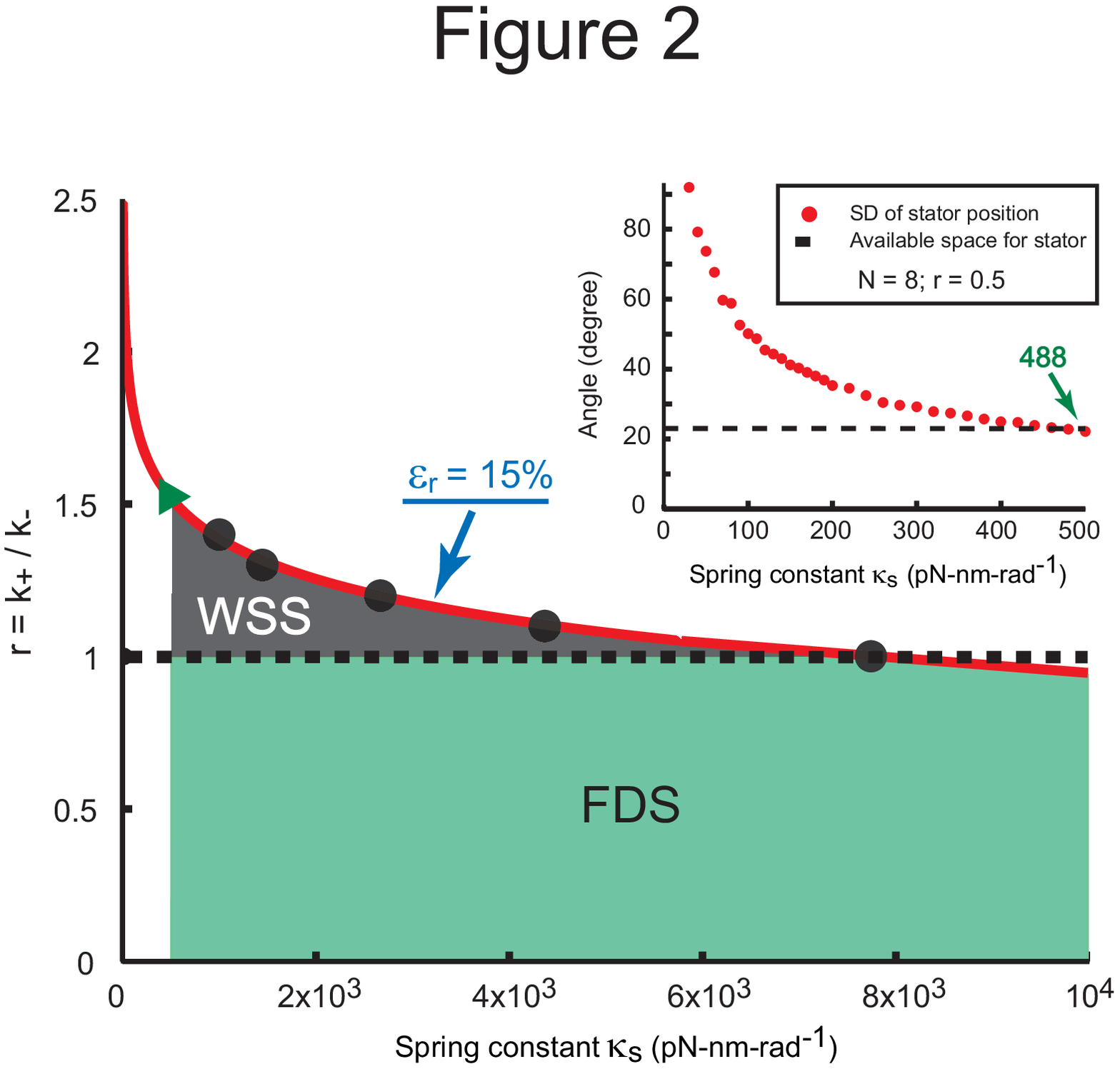}
   \end{center}
\end{figure}

\clearpage
\begin{figure}
   \begin{center}
      \includegraphics*[angle=0,width=5.25in]{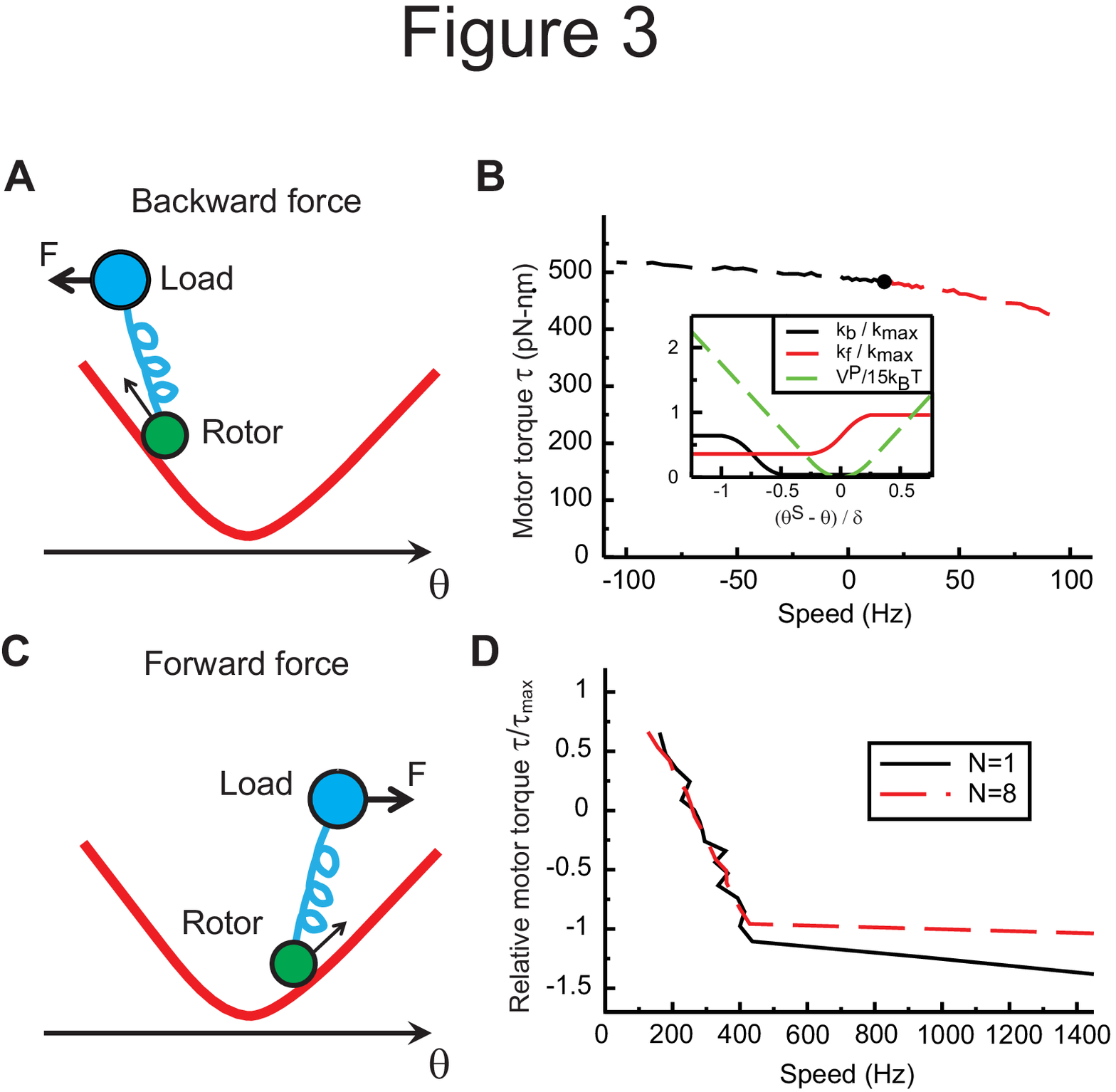}
   \end{center}
\end{figure}

\clearpage
\begin{figure}
   \begin{center}
      \includegraphics*[angle=0,width=5.in]{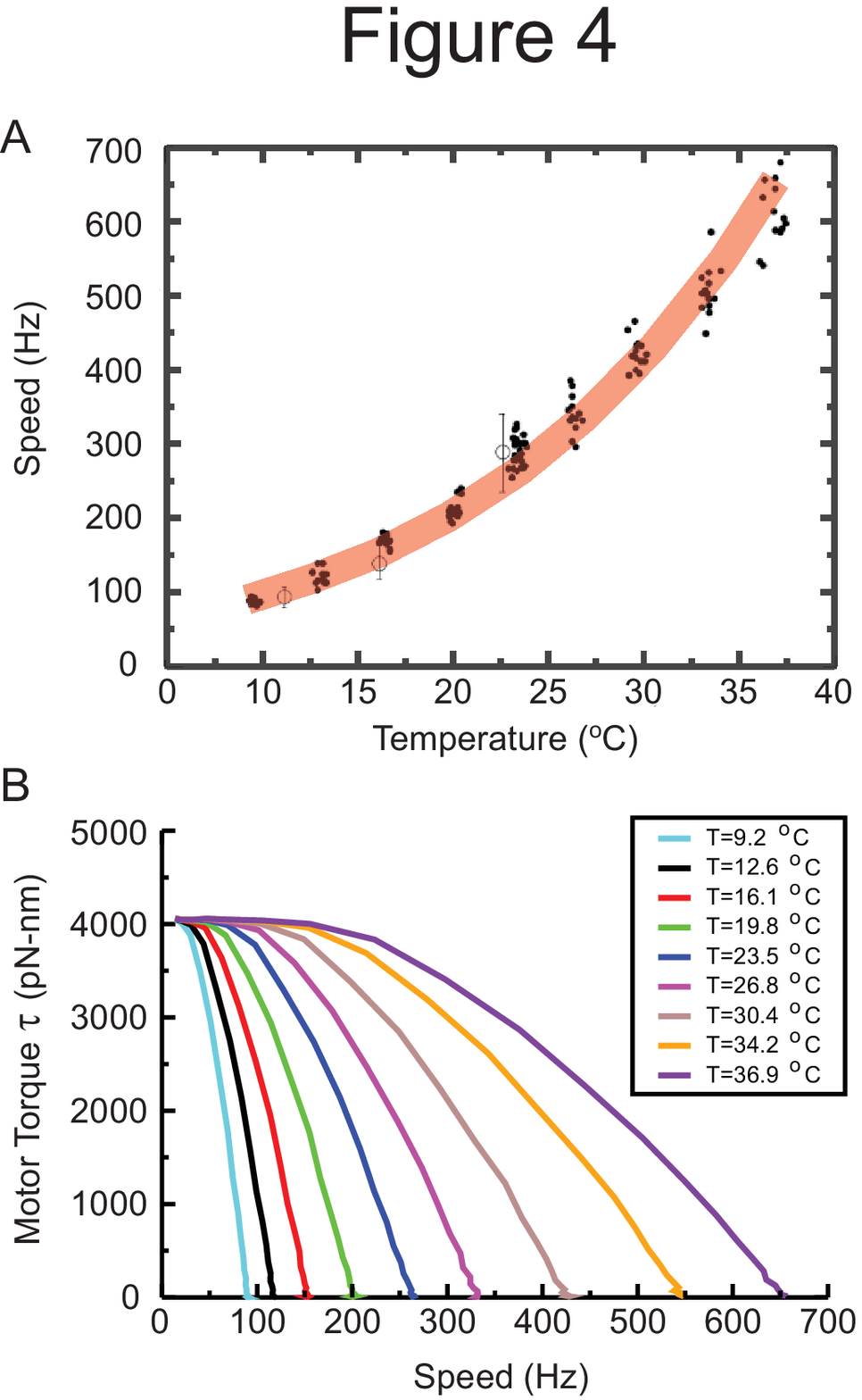}
   \end{center}
\end{figure}

\clearpage
\begin{figure}
   \begin{center}
      \includegraphics*[angle=0,width=3.5in]{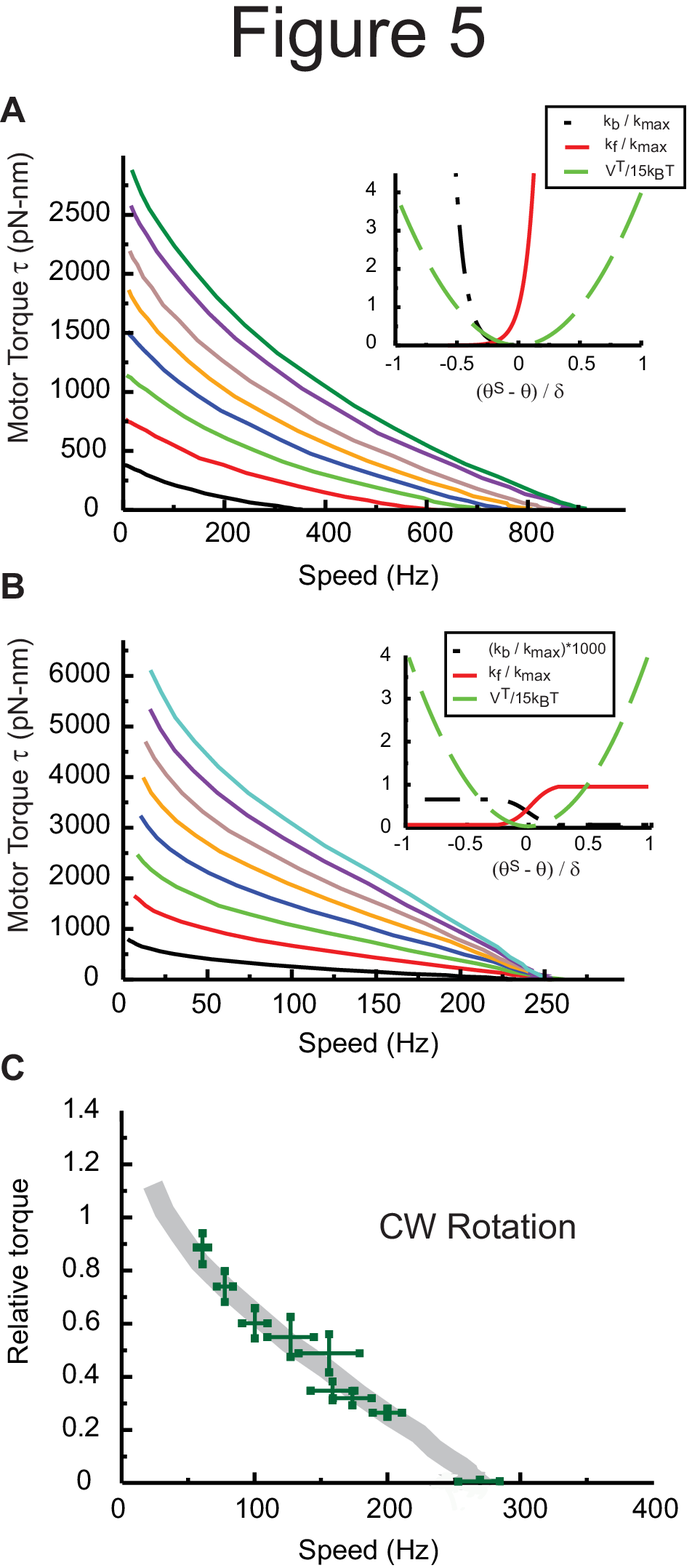}
   \end{center}
\end{figure}

\clearpage
\begin{figure}
   \begin{center}
      \includegraphics*[angle=0,width=3.5in]{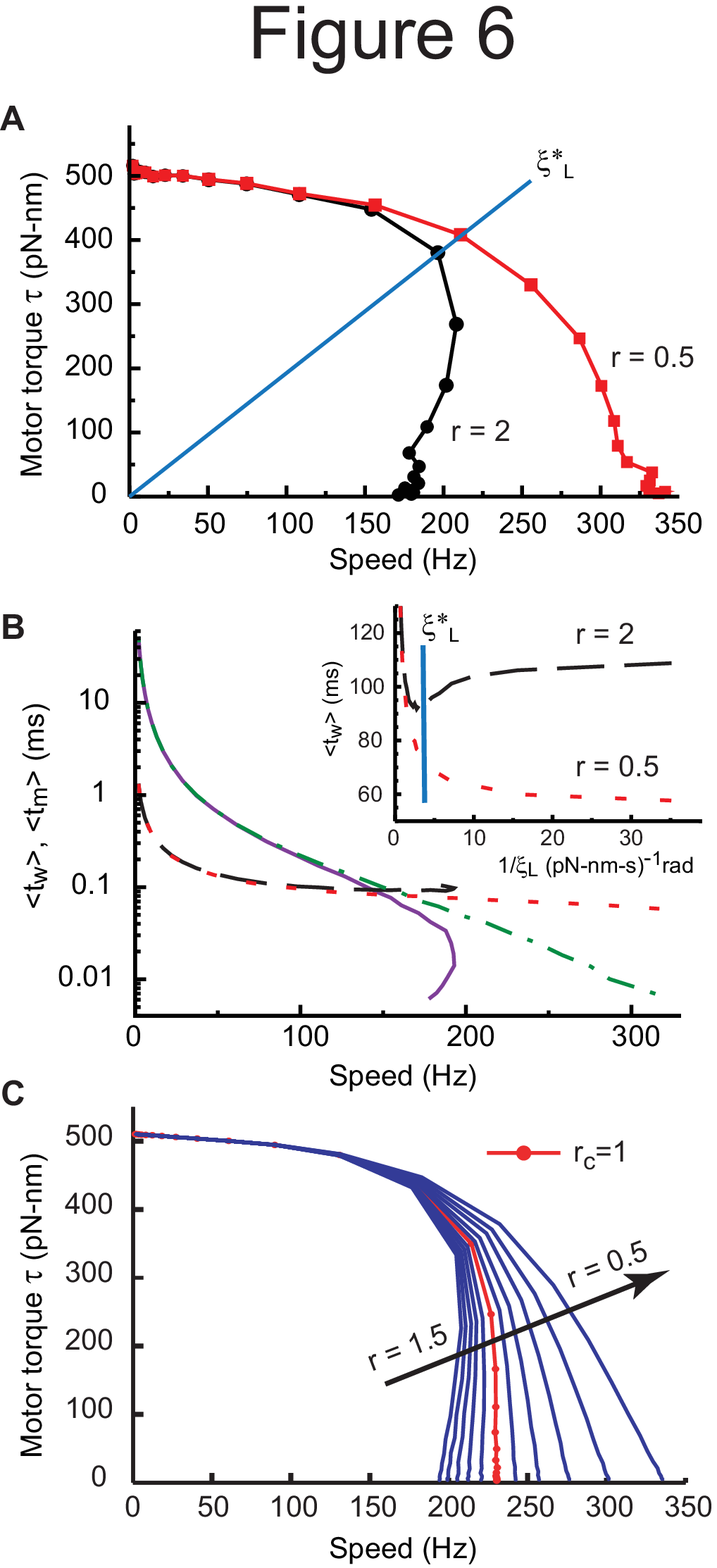}
   \end{center}
\end{figure}


\begin{thebibliography}{110}
\bibitem{note1}
In fact, the $\alpha$-helix spring cannot be stretched much more than its total length and the available space for stator fluctuation.
\end{thebibliography}
\end{document}


\begin{center}
{\Large \bf Supplementary Information\\}
\vspace{0.5cm}
{\large \bf
The Effects of Stator Compliance, Backs Steps, Temperature, and Clockwise Rotation on the Torque-Speed Curve of Bacterial Flagellar Motor.
\\}
\vspace{0.5cm}
Giovanni Meacci, Ganhui Lan, and Yuhai Tu\\
\vspace{0.3cm} IBM T. J. Watson Research Center\\ P.O. Box 218, Yorktown
Heights, NY 10598\\
\vspace{0.3cm}
\end{center}

\begin{flushleft}
\Large{\bf {Contents}}
\end{flushleft}
\begin{itemize}
\item
\large{$\alpha$-helix spring constant estimation}
\item
Stator spring effects at high and low load, and model parameter study
\item
Comments on an alternative models
\item
Rotation under external torque for N$>$1 and effects of backward jumping probability
\item
Force scale and stator conformational change in the jumping rate
\end{itemize}
\clearpage

\section*{\Large{Estimation of the $\alpha$-helix spring constant}}

In bacterial flagella motor, the stator is anchored onto the peptidoglycan cell wall through a short segment of $\alpha$-helix. When the motor rotates and the stator interacts with the rotor, the $\alpha$-helix exerts force that generates a torque to anchor the stator around its position. This physical process is abstracted as a stator spring in some models~\citep{Bai_et_al09}, and we have systematically studied the effect of this additional elastic element to flagella motor operation in the main text of this paper. In this supplemental section, we aimed to analyze the magnitude of the stator spring constant, which provides important insight to the physical limitation of determining model parameters.

Under most biophysical conditions, $\alpha$-helix can be regarded as an elastic rod. In general, $\alpha$-helix can bear bending, twisting and stretching deformations. And the elasticity of $\alpha$-helix has been theoretically studied~\citep{seungho2005elasticity}. The persistence length $l_p$ of an $\alpha$-helix rod, which determines the stiffness of a linear object against deformation, has been estimated to be $100$nm. In reality, the 3-dimensional deformation of the stator anchoring $\alpha$-helix could be complicated. A reasonable simplification is to assume that the helix only bends within the rotor plane. This assumption enables us to estimate the lower limit of the stator spring constant, which is also the key point with most debates among different theoretical models.

\begin{figure}
   \begin{center}
      \includegraphics*[angle=0,width=5.25in]{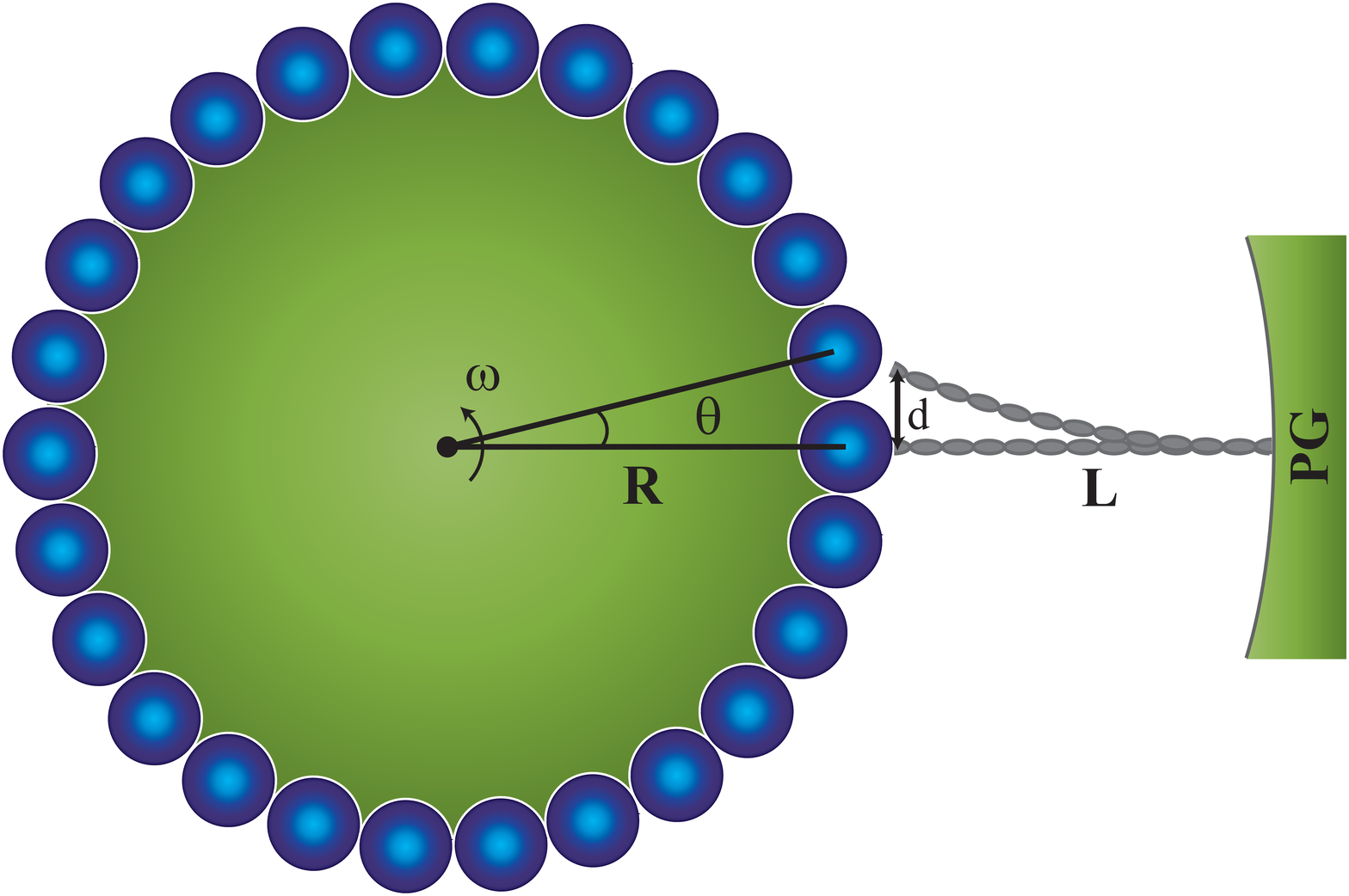}
      \caption{\small{$2$-dimensional flagellar motor geometry. The 26 blue dots represent the 26 FliG proteins on the rotor of radius R, rotated of an angle $\theta$ at velocity $\omega$. The gray rod represent the $\alpha$-helix of length L, and d its displacement tangential to the rotor. PG is the peptidoglycan cell wall. See also Figure 1 in reference \citep{Meacci_Tu09} for a schematic illustration of the rotor-stators spatial arrangement.}
}
      \label{fig:FigS1_AI}
   \end{center}
\end{figure}

For an $\alpha$-helix with length $L$, we can use the theory of linear elasticity to evaluate its deformation energy due to bending. According to the $2$-dimensional flagella motor geometry shown in Figure \ref{fig:FigS1_AI}, the elastic energy embedded in the $\alpha$-helix can be written as:

\begin{equation}
E_{\alpha}=\frac{1}{2}k_BT \int _L ds l_p \cdot \Delta^2(s)
\end{equation}

\noindent where $\Delta$ is the torsional angle along the bent $\alpha$-helix and $k_BT$ is the thermal energy. In the limit of small deformation, the torsional angle can be approximated using the end displacement of the $\alpha$-helix $d$: $\Delta \approx \frac{d/L}{L}$. And the energy becomes:

\begin{equation}
E_{\alpha}(d) \approx \frac{1}{2}k_BT \cdot L \cdot l_p \cdot (\frac{d/L}{L})^2 = \frac{1}{2}k_BT \cdot l_p \cdot \frac{d^2}{L^3}
\end{equation}

\noindent From Figure \ref{fig:FigS1_AI}, it is easy to obtain the geometrical relation: $d=R \cdot \theta$. So the energy contribution from stator spring in the rotor rotational coordinate is:

\begin{equation}
E_{\alpha}(\theta) \approx \frac{1}{2}k_BT \cdot l_p \cdot \frac{R^2}{L^3} \cdot \theta ^2
\end{equation}

\noindent And the stator spring constant $\kappa_S$ is:

\begin{equation}
\kappa_S = \frac{\partial ^2 E_{\alpha}(\theta)}{\partial \theta ^2} \approx k_BT \cdot l_p \cdot \frac{R^2}{L^3}
\end{equation}

\noindent A bacterial flagella motor rotor is about $R=20$nm in radius, the length of the elastic $\alpha$-helix is about $L=7$nm, and under room temperature, $1k_BT=4.2$pN$\cdot$nm, which gives rise to an estimated value for the stator spring: $\kappa_S \approx 500$pN$\cdot$nm$\cdot$rad$^{-1}$.

\section*{Stator spring effects at high and low load and model parameters study}

In this supplementary section, we explore how the stator spring affects the general behavior of a flagella motor. We focus on the three following points:
\begin{itemize}
 \item Stator spring effects on flagella motor dynamics at high and low load for jumping rate ratio $r=2$.
 \item Robustness study of load free speed ratio $\omega_r$ as a function of the stator spring constant for different values of the stator diffusion constant.
 \item Independence of the speed near zero load on the stator number for a large interval of values of parameters model, such the stator
     diffusion constant and the the stator spring, when r=0.5.
\end{itemize}

\begin{figure}
   \begin{center}
      \includegraphics*[angle=0,width=5.25in]{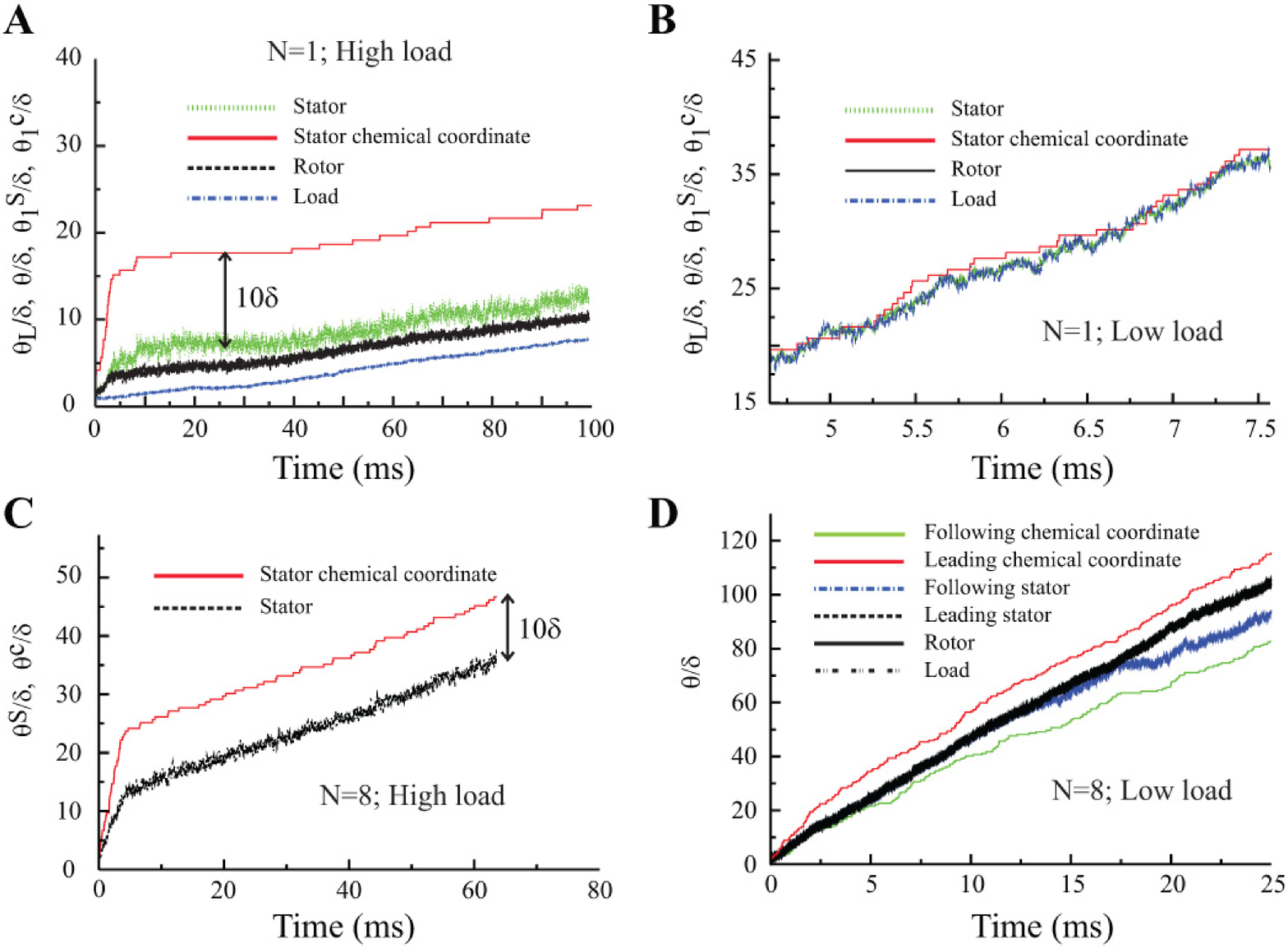}
      \caption{\small{The stator, rotor and load dynamics for the case r=2, $\kappa_S=$200pN$\cdot$nm$\cdot$rad$^{-1}$, $D_S=$500rad$^2$/s.
       ({\bf A}) High load with N=1. The red solid line shows the chemical internal coordinate of the stator $\theta^{c}_1$ and the green dashed line the dynamics of the corresponding stator: internal chemical coordinate $\theta^{c}_1$ plus the (physical) angular distance of the stator from its stationary anchoring point on the cell wall $\theta_1$. The stretching of the stator spring is roughly $10 \delta$.  Black and blue dashed lines are the rotor and load angular coordinates respectively.
       ({\bf B}) Low load with N=1. The red solid line shows  the chemical internal coordinate of the stator $\theta^{c}_1$ and the green dashed line the dynamics of the corresponding stator. Black and blue dashed lines are the rotor and load angular coordinates respectively. The rest position $(\theta-\theta_L)_0$ is set to zero.
      ({\bf C}) One of the eight stator (N=8) at high load. The red solid line shows the chemical internal coordinate of the stator of one stator i, and the black dots the corresponding $\theta^{S}_i$. The stretching is roughly $10 \delta$.
      ({\bf D}) Low load with N=8. The red solid line shows the chemical internal coordinate of the leading stator and the black dots the dynamics of the corresponding stator. The stretching between the two is roughly $10 \delta$. The solid green line shows the internal coordinate of one of the following stators and the blue dashed-dotted line the corresponding stator. Black solid line and black semi-dashed line, indistinguishable in the figure, represent the rotor and the load respectively.
      }
      }
      \label{fig:FigS2_AI}
   \end{center}
\end{figure}

Figure \ref{fig:FigS2_AI} shows typical cases of the stator dynamics at high and low load for different number of stators ($N$) with presence of stator springs. The stator spring constant is chosen to be $\kappa_S=200$pN$\cdot$nm$\cdot$rad$^{-1}$, in line with the value used in ref. \citep{Bai_et_al09}. We also choose the stator drag coefficient $\xi_S$ to make corresponding diffusion constant $D_S=k_B T/\xi_S=500 rad ^{2}s^{-1}$, the same as used in \citep{Bai_et_al09}, which is half of the value used in Yuan and Berg's paper \citep{Yuan_Berg08}. Figure \ref{fig:FigS2_AI}A presents the stator dynamics of $N=1$ at high load. The initial angular positions for the load and the rotor are set to be the same. At the beginning of the motion, the load is stretched for a transient time of $\sim 5$ms, after which the averaged speed of the internal coordinate becomes the same as the averaged stator speed, with a displacement of roughly $10 \delta$ between the stator position ($\theta_{1}$) and the stator anchoring point ($\theta^{c}_{1}$), where $\delta$ is size of a FliG unit. The transient time and the stator spring stretching both depend on the load and the stator spring constant. The results reported in the main text are all computed after the system reaches steady state.

Figure \ref{fig:FigS2_AI}B shows the behavior of single-stator motor at low load (high speed) regime. From the simulated trajectory, it is easy to see that the stator motor motion exhibits significant fluctuation, although the stator spring only get stretched less than $1 \sim 2 \delta$. In this regime, the load does not show noticeable displacement with respect to the rotor.

Increasing number of stators influences motor dynamics at high and low load regimes in different ways. At high load, the qualitative behavior for a $8$-stator motor is the same as for a single-stator motor. Each spring is in average stretched $\sim 10 \delta$ in the same direction as the motor rotation. Figure \ref{fig:FigS2_AI}C shows the dynamics of one of the eight stators at high load. All other stators behave the same (data not shown). The synchronized property does not exist for motor motion under low load. Each stator is stretched differently and there exists faster stator, i.e. the stator that jump more frequently, who leads the motor of the rotor (as shown in Figure \ref{fig:FigS2_AI}D). Also the directions of stators displacements now are different with some stator spring stretched in the same direction as the motor rotation and same others in the opposite one. In both directions, the stretching between the rotor and the stator's chemical coordinate (and physical position) can be as large as many folds of $\delta$. To keep the stator fluctuation within $2 \delta$ under the entire motor operating spectrum, the stator spring has to be stiffer than $800$pN$\cdot$nm$\cdot$rad$^{-1}$, which is $4$ times larger than the maximum stator spring constant ($200$pN$\cdot$nm$\cdot$rad$^{-1}$) used in \citep{Bai_et_al09} to ensure the independence of the load free speed on the number of stators. In the main text, Figure 2, we further loosened the allowed limit for stator fluctuation, but the obtained stator spring constant ($\sim 500$pN$\cdot$nm$\cdot$rad$^{-1}$) would still cause significant ($>30\%$) deviation in the load free speeds at different number of stators.

\begin{figure}
   \begin{center}
      \includegraphics*[angle=0,width=4.in]{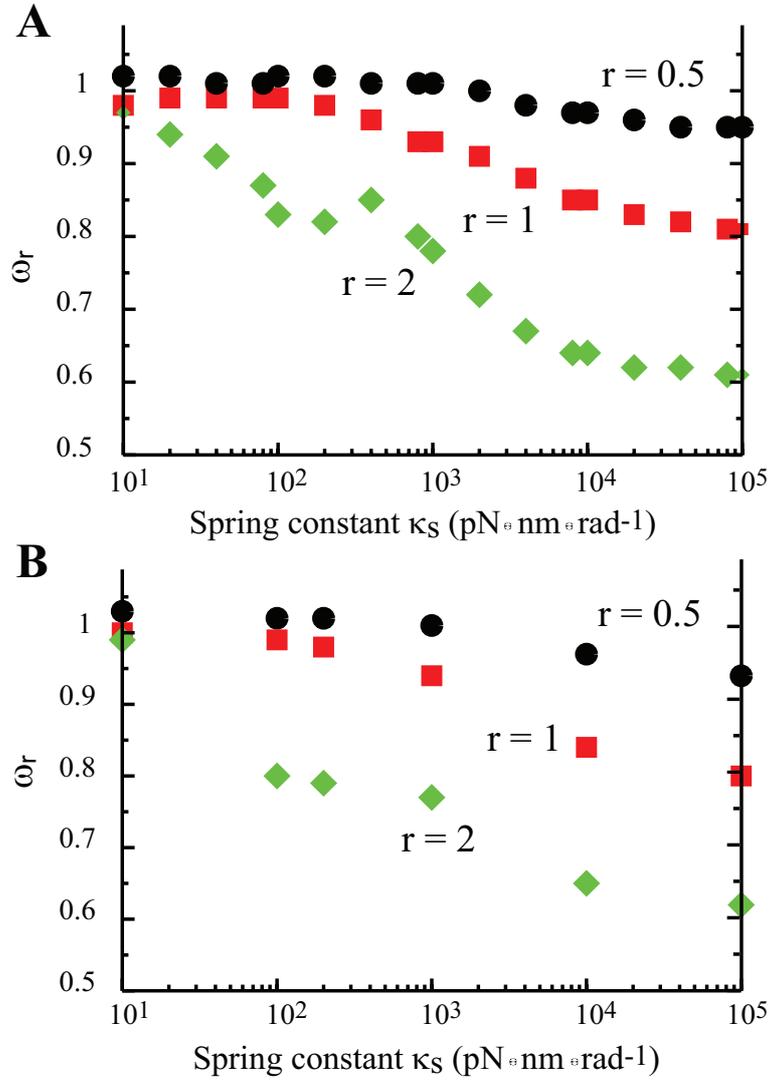}
      \caption{\small{({\bf A}) $\omega_r = \omega(8)/\omega(1)$ as a function of $\kappa_S$ with $D_{S}=50rad^{2}s^{-1}$. Black dots, red squares, and green diamonds, correspond to r=0.5, r=1, and r=2 respectively.
      ({\bf  B}) $\omega_r = \omega(8)/\omega(1)$ as a function of $\kappa_{S}$ with $D_{S}=5000rad^{2}s^{-1}$. Black dots, red squares, and green diamonds, correspond to r=0.5, r=1, and r=2 respectively.}
      }
      \label{fig:FigS3_AI}
   \end{center}
\end{figure}

In Figure \ref{fig:FigS3_AI}, we present the stator spring dependence of the load free speed ratio $\omega_r = \omega(8) / \omega(1)$ at different stator diffusion constant $D_S$. We evaluated the dependence at low diffusivity $D_{S}=50$rad$^{2}$s$^{-1}$ and high diffusivity $D_{S}=5000$rad$^{2}$s$^{-1}$. The dependence is very similar in these two extreme choices of $D_{S}$ and is also similar to the intermediate $D_{S}=500$rad$^{2}$s$^{-1}$ as shown in Figure 1B of the main text result.

\begin{figure}
   \begin{center}
      \includegraphics*[angle=0,width=5.25in]{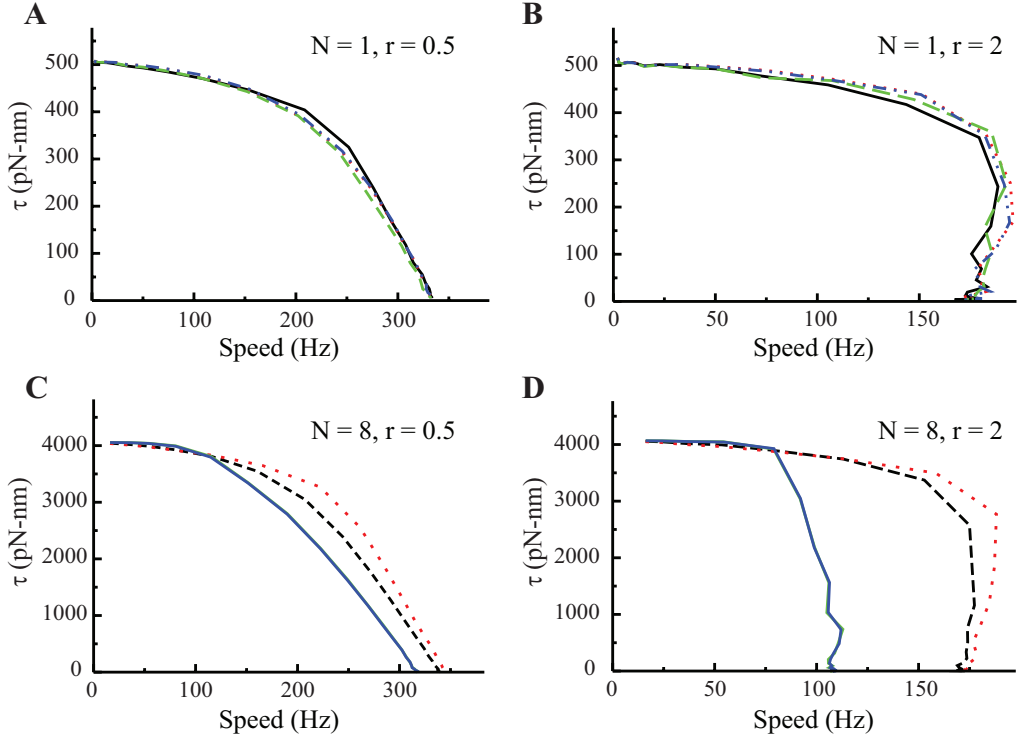}
      \caption{\small{({\bf A}) Torque-speed curves for four different combinations of $\kappa_S$ and $D_S$, when r=0.5 and N=1. Solid black line $\kappa_S=10$pN$\cdot$nm$\cdot$rad$^{-1}$, $D_S=5000$. Dotted red line $\kappa_S=10^5$pN$\cdot$nm$\cdot$rad$^{-1}$, $D_S=5000$. Dashed green line $\kappa_S=10$pN$\cdot$nm$\cdot$rad$^{-1}$, $D_S=50$. Dashed-dotted blue line $\kappa_S=10^5$pN$\cdot$nm$\cdot$rad$^{-1}$, $D_S=50$.
      ({\bf B}) Torque-speed curves for four different combinations of $\kappa_S$ and $D_S$, when r=2 and N=1. Solid black line $\kappa_S=10$pN$\cdot$nm$\cdot$rad$^{-1}$, $D_S=5000rad^{2}s^{-1}$. Dotted red line $\kappa_S=10^5$pN$\cdot$nm$\cdot$rad$^{-1}$, $D_S=5000rad^{2}s^{-1}$. Dashed green line $\kappa_S=10$pN$\cdot$nm$\cdot$rad$^{-1}$, $D_S=50rad^{2}s^{-1}$. Dashed-dotted blue line $\kappa_S=10^5$pN$\cdot$nm$\cdot$rad$^{-1}$, $D_S=50rad^{2}s^{-1}$
      ({\bf C}) Torque-speed curves for four different combinations of $\kappa_S$ and $D_S$, when r=0.5 and N=8. Dashed black line $\kappa_S=10$pN$\cdot$nm$\cdot$rad$^{-1}$, $D_S=50rad^{2}s^{-1}$. Dotted red line $\kappa_S=10$pN$\cdot$nm$\cdot$rad$^{-1}$, $D_S=5000rad^{2}s^{-1}$. Solid green line $\kappa_S=10^5$pN$\cdot$nm$\cdot$rad$^{-1}$, $D_S=50rad^{2}s^{-1}$. Solid blue line $\kappa_S=10^5$pN$\cdot$nm$\cdot$rad$^{-1}$, $D_S=5000rad^{2}s^{-1}$.
      ({\bf D}) Torque-speed curves for four different combinations of $\kappa_S$ and $D_S$, when r=2 and N=8. Dashed black line $\kappa_S=10$pN$\cdot$nm$\cdot$rad$^{-1}$, $D_S=50rad^{2}s^{-1}$. Dotted red line $\kappa_S=10$pN$\cdot$nm$\cdot$rad$^{-1}$, $D_S=5000rad^{2}s^{-1}$. Solid green line $\kappa_S=10^4$pN$\cdot$nm$\cdot$rad$^{-1}$, $D_S=50rad^{2}s^{-1}$. Solid blue line $\kappa_S=10^5$pN$\cdot$nm$\cdot$rad$^{-1}$, $D_S=5000rad^{2}s^{-1}$.}
      \label{fig:FigS4_AI}
      }
   \end{center}
\end{figure}

In the case of $r=0.5$, all torque-speed curves, with $N$ from $1$ to $8$, collapse near zero load for all choices of stator spring constant $\kappa_S$ crossing from $10$pN$\cdot$nm$\cdot$rad$^{-1}$ to $10^5$pN$\cdot$nm$\cdot$rad$^{-1}$, and of stator diffusion constant $D_{S}$ crossing from $50$rad$^{2}$s$^{-1}$ to $5000$rad$^{2}$s$^{-1}$. This behavior is shown in Figure \ref{fig:FigS4_AI}A and \ref{fig:FigS4_AI}C. On the contrary, the torque-speed curves collapse only for small values of the stator spring constant, e.g. $\kappa_S=10$pN$\cdot$nm$\cdot$rad$^{-1}$, when setting ratio of jumping rates to be $r=2$ (Figure \ref{fig:FigS4_AI}B and \ref{fig:FigS4_AI}D). It is also easy to see from Figure \ref{fig:FigS4_AI}C and \ref{fig:FigS4_AI}D that increasing stator spring constant $\kappa_S$ decreases the load free speed of the 8-stator motor, and the magnitude of decreasing is proportionally bigger for $r=2$ than for $r=0.5$.

\section*{Comment on an alternative models}

In this supplemental section, we will discuss the aspects of the backward motion and the stator spring effect for alternative models proposed in references~\citep{XBBO06,Bai_et_al09,vanAlbada2009}.

When external load exceeds the stall-torque value, stator would be forced to move backwards which overpowers the stator-rotor interaction. In the models presented in refs. \citep{XBBO06,Bai_et_al09} (Figure 4B and Figure 1 in the referenced papers), sharp peaks in the interaction potential were introduced to prevent thermal fluctuations from taking the motor down to the back side of the potential field. This setup inevitably results in discontinuity in the torque value for motor backward motion (a small barrier for backward rotation), which is inconsistent with the experimental observations for {\it E. coli} flagellar motor. Moreover, according to this potential setup, after the motor moves across the barrier, the required external torque to maintain backward motion would sharply decrease, so at near stall torque regime, strong fluctuation on the torque values is expected. We are expecting the same behavior also in the model with periodic potential used in ref. \citep{vanAlbada2009}. However, this prediction has not been observed in any experiments.

Another discrepancy between Bai's model and experiments is the stator spring effect on motor dynamics. As pointed out in the main text and in previous supplemental sections, for the chemical jumping rate ratio $r=2$, the stator spring constant $\kappa_S \le 200$pN$\cdot$nm$\cdot$rad$^{-1}$~\citep{XBBO06,Bai_et_al09} gives rise to a $30\%$ deviation in the load free speeds between single-stator and 8-stator motors. Another unphysical consequence from Bai's model is the large fluctuation of stator position due to soft spring (Figure \ref{fig:FigS2_AI} and Figure 2 in the main text). And the fluctuation could be more than $2$ FliG unit size, which could lead to physical contact of neighboring stators during rotation.

\section*{Rotation under external torque for $N>1$ and effects of backward jumping probability}

In this supplemental section, we extended our study of flagellar motor rotation under external torque from single-stator ($N=1$) to multi-stator ($N>1$). The stator-rotor potential and the jumping rates are defined in the main text.

\begin{figure}
   \begin{center}
      \includegraphics*[angle=0,width=4.25in]{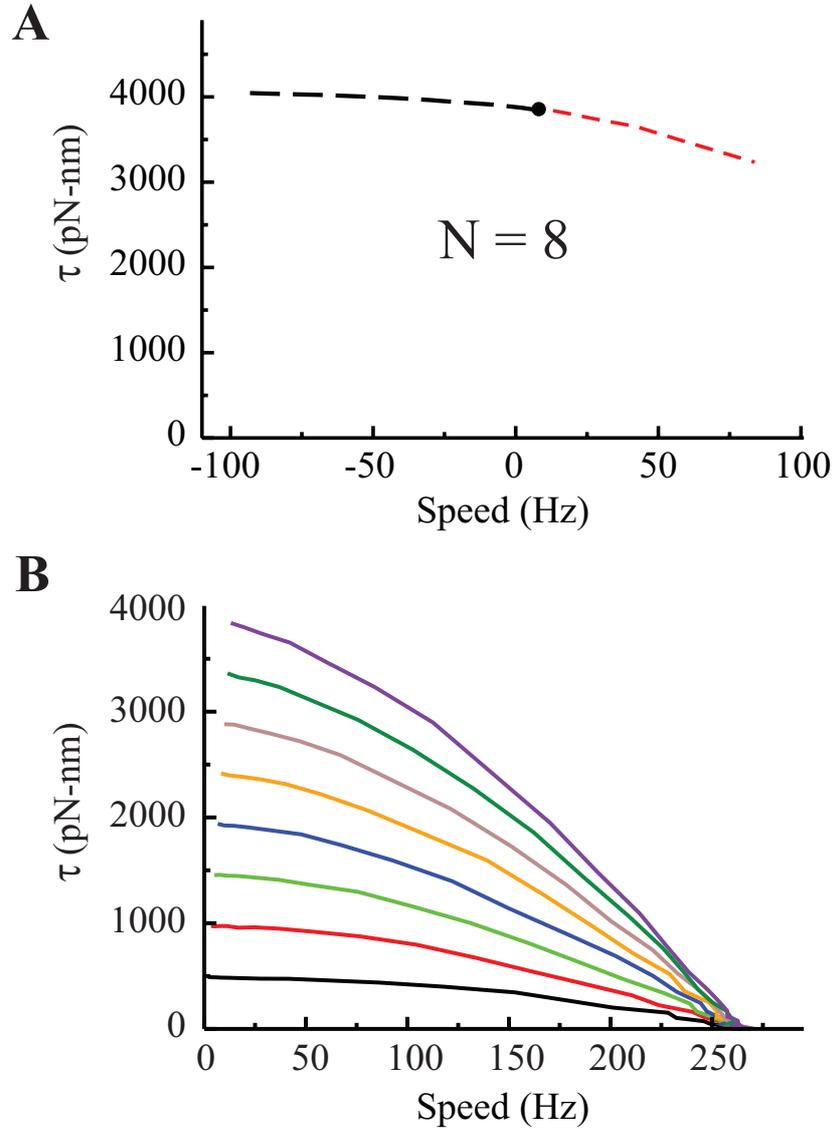}
      \caption{\small{({\bf A}) Torque-speed curve for N=8 near stall. Starting from the black dot, the dashed black and red lines represent the cases when the applied external torque has respectively the opposite and the same sign as the stator generated torque.
      ({\bf B}) Torque-speed curves, from N=1 to N=8, for the case with backward jumping probability. Parameter values as in Figure 3 of the main text.}
      }
      \label{fig:FigS5_AI}
   \end{center}
\end{figure}

We first confirmed that the toque-speed curve is still linear with speed up to $100$Hz in both rotational directions. Figure \ref{fig:FigS5_AI}A shows an example for $N=8$.

We then computed the entire toque-speed relation in normal motor operative range when the backward jumping probability is introduced. Results show that the major characteristic shape of the torque-speed curves remains the same, especially the independence of the maximum speed on number of stators is preserved (Figure \ref{fig:FigS5_AI}B).

It is worth mentioning that sometimes motors break when large external torque is applied \citep{Turner96}. Some motors break catastrophically: the cell suddenly spins at higher speed and then stop spinning when the external torque is turned on. However, most motors broke progressively. In this case, motor can be partially or completely recovered. A plausible explanation for these behaviors is that the catastrophic breaks disconnect the drive shaft, whereas progressive breaks disassemble the torque generating units, which can be replaced by the cell on time scale of several minutes.

\section*{Force scale and stator conformational change in the jumping rate}

The independence of the speed near zero load on the stator number obtained with parabolic potential and jumping rates given by equations (14) and (15) in the main text does not rely on the specific value of the force scale $\tau_c=E_0/\delta\theta_l$ used in the expression of the jumping rate. Figure S6A shows the case with $\tau_c \approx 0.2 \tau_0$.

A lower bound for $\delta \theta _l$, which can be interpreted as the size of the conformational change of the stator's transition state before stepping, has been estimated to be 0.5$^ \circ$ (main text) from the maximum torque $\tau_0$ and the jumping rate ratio $r$ , which correspond to roughly $0.05 \delta$. Figure S6B shows torque-speed curves of single-stator motor for different values of $\delta \theta _l$, starting from the lower value. For $0.05\delta < \delta \theta _l < 0.3\delta$ all curves fall in the experimental observed values of rotational speed and exerting torque.

\bibliographystyle{plainnat}

\clearpage

\clearpage

\clearpage
\clearpage

\clearpage
\clearpage
\begin{figure}
   \begin{center}
      \includegraphics*[angle=0,width=4in]{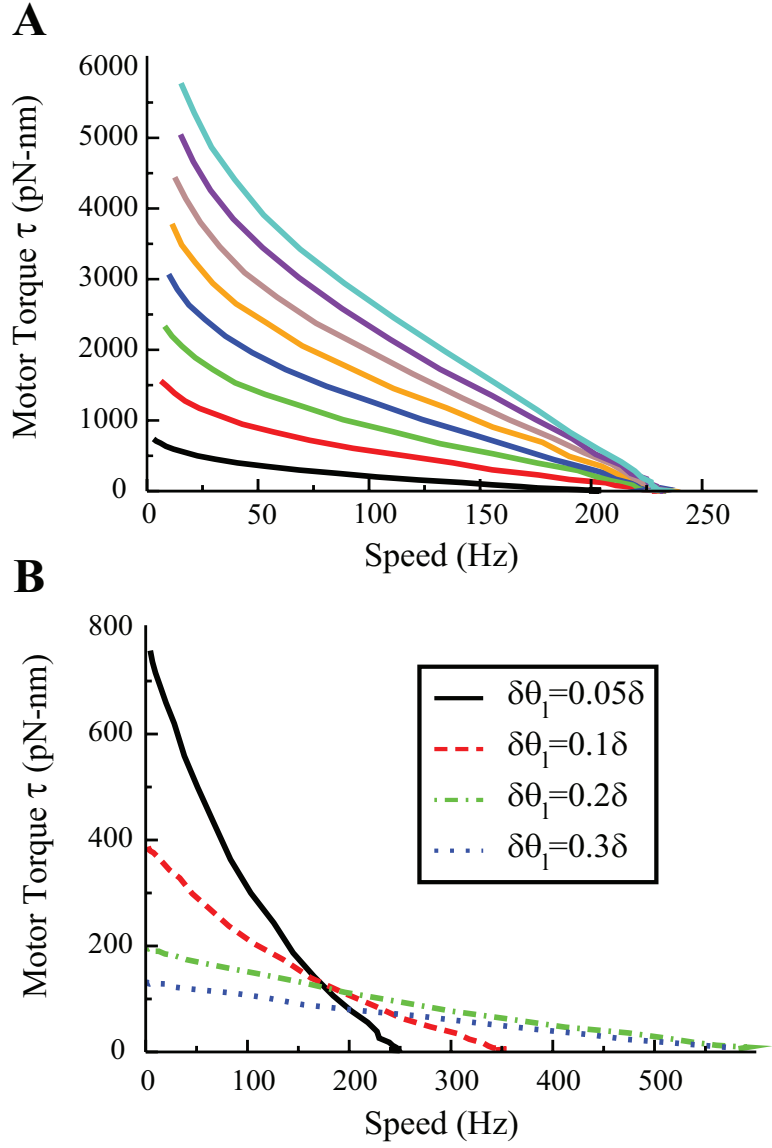}
      \caption{({\bf A}) The torque-speed ($\tau-\omega$) curves for different number of stators. N increases, from N=1 to N=8, starting from the black curve at low torque value to the dark green at high torque value. Parameter values are: $k_0=30000s^{-1}$, $\kappa_S=10^{4}$pN$\cdot$nm$\cdot$rad$^{-1}$, $D_S=500rad^2s^-1$, $k_r=0.01$, $\delta\theta_l=0.1\delta$,
      $E_c=0.2 \tau_0 \delta\theta_l$, i.e. $\tau_c \approx 0.2 \tau_0$, and with parabolic potential.
      ({\bf B}) The torque-speed ($\tau-\omega$) curves for different values of $\delta \theta_l$. N=1, $\kappa_S=10^{4}$pN$\cdot$nm$\cdot$rad$^{-1}$, $D_S=500rad^2s^-1$, $k_r=0.01$ and with parabolic potential. The jumping rates are given by equations (14) and (15) in the main text. Solid black line $\delta\theta_l=0.05\delta$, red dashed line $\delta\theta_l=0.1\delta$, green dashed-dotted line $\delta\theta_l=0.2\delta$, and blue dotted line $\delta\theta_l=0.3\delta$.}
      \label{fig:FigS6_AI}
   \end{center}
\end{figure}
